\documentclass[11pt,aip,jap]{article}
\usepackage[utf8]{inputenc}
\usepackage{setspace}
\usepackage{cite}
\usepackage{graphicx}
\usepackage{float}
\usepackage{amsmath}
\usepackage{authblk}
\usepackage{blindtext}
\usepackage{color,soul}
\usepackage{gensymb}
\usepackage{booktabs,chemformula}
\usepackage{geometry}
\usepackage{xcolor}
\usepackage{soul}
\providecommand{\keywords}[1]
{
  \small	
  \textbf{\textit{Keywords---}} #1
}

\title{Interface controlled thermal properties of ultra-thin chalcogenide-based phase change memory devices}
\author[1]{Kiumars Aryana}
\author[1]{John T. Gaskins}
\author[2]{Joyeeta Nag}
\author[2]{Derek A. Stewart}
\author[2]{Zhaoqiang Bai}
\author[3]{Saikat Mukhopadhyay}
\author[2]{John C. Read}
\author[1]{David H. Olson}
\author[4]{Eric R. Hoglund}
\author[4]{James M. Howe}
\author[5]{Ashutosh Giri}
\author[2]{Michael K. Grobis}
\author[1,4,6]{Patrick E. Hopkins\thanks{Corresponding Author: phopkins@virginia.edu}}
\affil[1]{Department of Mechanical and Aerospace Engineering, University of Virginia, Charlottesville, Virginia 22904, USA}
\affil[2]{Western Digital Corporation, San Jose, CA 95119, USA}
\affil[3]{NRC Research Associate at Naval Research Laboratory, Washington, DC 20375, USA}
\affil[4]{Department of Materials Science and Engineering, University of Virginia, Charlottesville, Virginia 22904, USA}
\affil[5]{Department of Mechanical, Industrial and Systems Engineering, University of Rhode Island, Kingston, RI 02881, USA}
\affil[6]{Department of Physics, University of Virginia, Charlottesville, Virginia 22904, USA}

\begin{document}
\date{}
\maketitle

\begin{abstract}
    Phase change memory (PCM) is a rapidly growing technology that not only offers advancements in storage-class memories but also enables in-memory data storage and processing towards overcoming the von Neumann bottleneck. In PCMs, the primary mechanism for data storage is thermal excitation. However, there is a limited body of research regarding the thermal properties of PCMs at length scales close to the memory cell dimension and, thus, the impact of interfaces on PCM operation is unknown. Our work presents a new paradigm to manage thermal transport in memory cells by manipulating the interfacial thermal resistance between the phase change unit and the electrodes without incorporating additional insulating layers. Experimental measurements show a substantial change in thermal boundary resistance as GST transitions from one crystallographic structure (cubic) to another (hexagonal) and as the thickness of tungsten contacts is reduced from five to two nanometers. Simulations reveal that interfacial resistance between the phase change unit and its adjacent layer can reduce the reset current for 20 and 120 nm diameter devices by up to $\sim$40\% and $\sim$50\%, respectively. The resultant phase-dependent and geometric effects on thermal boundary resistance dictate that the effective thermal conductivity of the phase change unit can be reduced by a factor of four, presenting a new opportunity to reduce operating currents in PCMs.
\end{abstract}

 \\

The growing demands for higher capacity memory devices and burgeoning data-intensive applications, such as artificial intelligence, have intensified efforts to beat the von Neumann computing bottleneck that separates processing from the storage unit. A promising alternative for transistor-based non-volatile memory devices is an emerging technology known as phase change memory (PCM), which offers prospective gains in speed, device lifetime, and storage capacity, as well as in-memory storage and computing capabilities \cite{wong2010phase, zhang2019designing}. The most widely used phase change material, germanium antimony telluride (GST), possesses a high electrical contrast between its amorphous and crystalline structure, as well as sub-nanosecond switching times \cite{simpson2011interfacial, rao2017reducing}. This class of phase change materials can quickly switch phase between amorphous and crystalline states upon controlled thermal excitation. In PCMs, the transition from amorphous to crystalline and crystalline to amorphous are commonly referred to as set and reset, respectively. In devices utilizing phase change units, thermal transport plays a pivotal role as it dictates the efficiency of the set/reset process and overall power consumption.

One of the major limitations in PCM devices is their high operating current, leading to excessive power consumption \cite{loke2012breaking}. In order to mitigate thermal leakage during programming, Kim \textit{et al}. \cite{kim2008fullerene} used a thermal barrier (2-20 nm of C$_{60}$) to insulate the GST from directly contacting the electrode, showing a factor of three reduction in their set current ($I_{set}$). Although a lower power consumption in their device architecture offered performance gains, the relatively large thickness of the thermal barrier introduced additional resistance, decreased bit density, and provided an additional source of degradation for the PCM over time. Later, Ahn \textit{et al}. \cite{ahn2015energy} proposed a much thinner insulating layer by using a single sheet of graphene (thickness $<$1 nm) as a thermal barrier to isolate the heat inside the PCM cell and showed that the $I_{reset}$ was reduced by $40\%$ compared to the cells without a graphene barrier. More recently, superlattice phase change memories have received a great deal of attention due to their unique capabilities offering lower power consumption, faster programming rate, higher retention time, and lower noise and drift in electrical resistance \cite{simpson2011interfacial, shen2019thermal, ding2019phase, saitochalcogenide}. Although earlier superlattice PCMs consisted of GeTe/Sb$_2$Te$_3$ alternating stacks, it was soon realized that this configuration tends to intermix and transform into bulk GST at high annealing temperatures \cite{momand2015interface}. Nonetheless, the idea of superlattice PCMs inspired researchers to look for alternative material configurations. Very recently, Shen \textit{et al.} \cite{shen2019thermal} and Ding \textit{et al.} \cite{ding2019phase} showed that superlattice PCMs with TiTe$_2$/Sb$_2$Te$_3$ layers have superior properties compared to bulk GST. Despite the fact that in superlattice PCMs the interface is an integral component in the performance of these devices, its effect on the overall thermal transport is heretofore unknown and unstudied. With all these previous works in mind, we are prompted to experimentally investigate the effect of interfacial thermal resistance on the performance of PCM devices. The selected materials for this study are amongst those that are widely used in PCM devices: Ge$_2$Sb$_2$Te$_4$ as a phase change unit, tungsten (W) as an electrode, and silicon dioxide (SiO$_2$) and silicon nitride (SiN$_x$) as the insulating separators used to confine heat and current within the cell. Our work focuses on identifying the critical parameters that influence thermal transport as the length scale of the phase change unit approaches that of energy carriers' mean free paths. We assess the effect of GST film thickness on thermal transport across various phase transitions and determine the minimum thickness before which thermal transport transitions into a ballistic regime. A pictorial representation of the configuration of layers used in this study is given in Fig. \ref{fig:Fig_01} (a) along with the corresponding transmission electron microscopy (TEM) images for amorphous (\textit{a}-GST) and hexagonal (h-GST) phases. 

To date, the majority of studies investigating thermal transport in GST were performed on layers with thicknesses on the order of 200 nm \cite{lyeo2006thermal, risk2009thermal,li2011grain, scott2020thermal}. However, as Xiong \textit{et al.} \cite {xiong2011low} demonstrated, in order to decrease power consumption and further the economic benefits of PCM memory devices, the thickness of GST layers should be on the order of 10 nm. In this respect, Kim \textit{et al.} devised an operational PCM device with cell dimensions as small as 7.5 nm $\times$ 17 nm \cite{kim2010high}. In general, as the length scale of materials and interconnects in PCM components shrink to dimensions less than energy carrier mean free paths, a number of additional mechanisms, such as electron tunneling \cite {raoux2014phase, xiong2013self, burr2014access} and thermal boundary resistances \cite{swartz1989thermal, hopkins2013thermal, scott2018thermal}, may impact the performance of these devices \cite{bozorg2010thermal}. In this paper, we present evidence of ballistic transport of energy carriers across the electrodes in a confined cell geometry as the characteristic length of the device is decreased to less than the mean free path of the carriers. We show that for tungsten electrodes there is a lower limit for the thickness of GST in memory cells before thermal transport transitions from a diffusive to a ballistic regime. Thus, our work uses this knowledge of carrier dynamics to experimentally identify an optimal thickness of phase change material based on a balance of thermal conductivity and crystallographic-phase-dependent thermal boundary conductances (TBC) in order to improve memory device performance.

Here, in contrast with previous studies that were primarily focused on introducing additional layers between the electrode and GST to confine heat in the memory cell, we focus on the interfacial thermal resistance and thermal properties of the layers in contact with GST. We show that, by intentionally engineering the phase and thickness of the phase change unit, the overall thermal resistance can be substantially increased, causing decreases in requirements for set/reset currents, without incorporating additional layers as a thermal barrier. Although the results presented here are for commonly used materials in PCMs such as GST and W, we demonstrate that, through manipulation of the interfacial resistance between the phase change unit and the adjacent layer, the predicted reset current can be reduced by up to 40$\%$ and 50$\%$ for devices with lateral size of 20 and 120 nm in diameters, respectively. Our work highlights the importance of engineering interfaces to allow for devices with increased performance.

\section*{Results}
We measure the thermal transport properties of the GST thin films, deposited via magnetron sputtering, using time-domain thermoreflectance (TDTR) in a two-tint configuration \cite{kang2008two}. The surface of the samples are coated with an 80 nm ruthenium transducer. The thermal model, which relates the thermoreflectivity of the transducer to the thermal properties of the underlying layers, requires knowledge of the volumetric heat capacity, film thickness, and thermal conductivity of each layer. The volumetric heat capacity for Ru, \textit{a}-GST, h-GST, and the Si substrate are assumed to be 2.96, 1.3, 1.4, and 1.64 MJ m$^{-3}$ K$^{-1}$, respectively \cite{furukawa1974critical,zalden2014specific}. The thermal conductivity of the Ru layer is determined to be 54 W m$^{-1}$ K$^{-1}$ via four point probe measurements and the thickness of each layer is confirmed via TEM.

\begin{figure}
\begin{center}
\includegraphics[scale = 1]{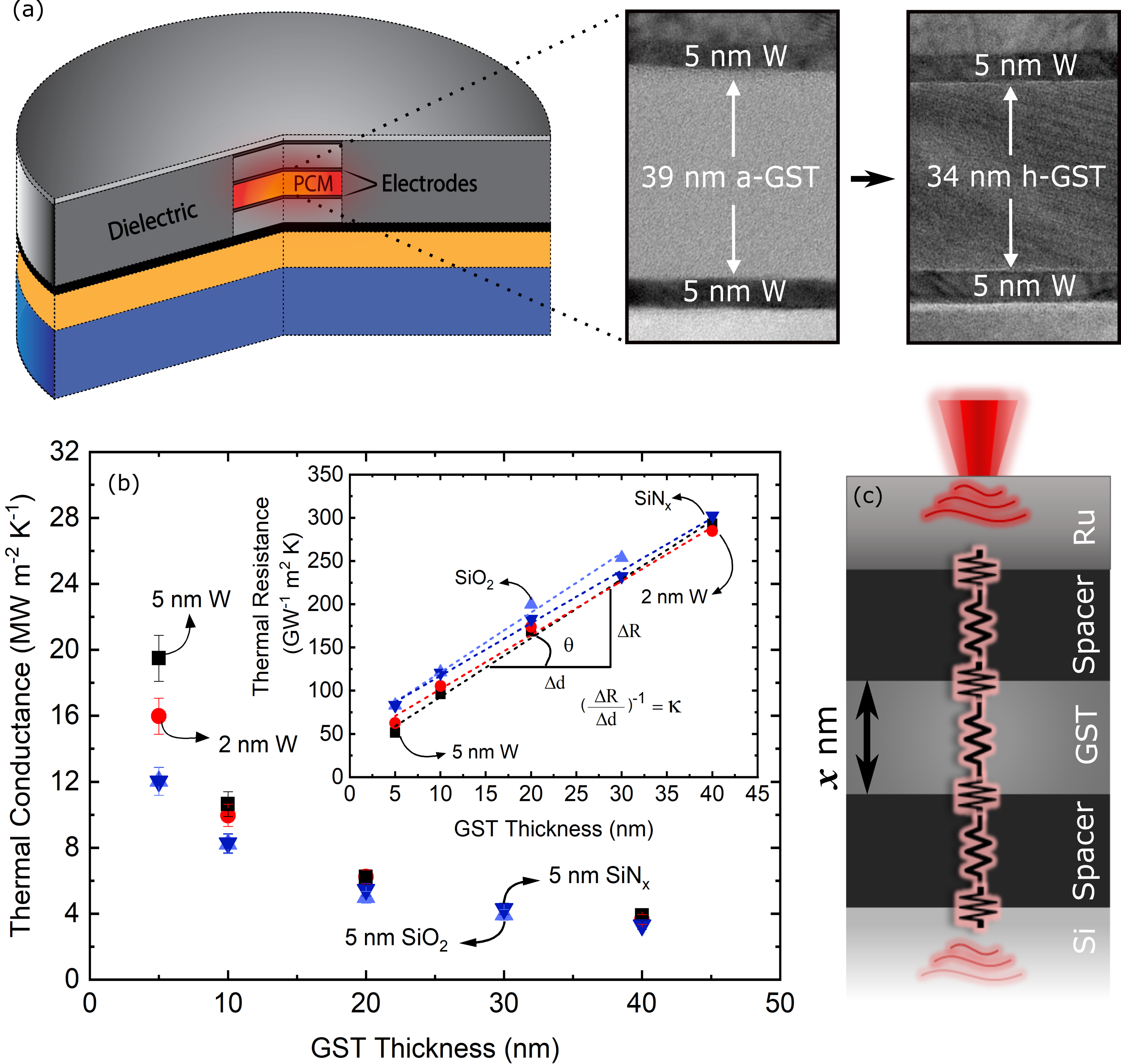}
\caption{(a) Schematic of a confined phase change memory cell along with corresponding TEMs for a 40 nm \textit{a}-GST and h-GST film sandwiched between 5 nm tungsten spacers, (b) thermal conductance across Ru/spacer/GST/spacer/Si for different spacer compositions as a function of GST thickness. The inset shows thermal resistance as a function of thickness where the inverse of the slope of the fitted line corresponds to the \textit{a}-GST thermal conductivity. The average thermal conductivity estimated for \textit{a}-GST is 0.15 $\pm$ 0.02 W m$^{-1}$ K$^{-1}$, and (c) a schematic of the thermal resistances in series for the multilayer configurations studied here. 
\label{fig:Fig_01}}
\end{center}
\end{figure}

\textbf{Room temperature thermal properties of \textit{a}-GST.}
We first investigate room temperature thermal properties in order to show that the effect of spacers on thermal conductance is only appreciable for \textit{a}-GST thicknesses less than 10 nm. For films as thin as these, it is often instructive to idealize material stacks as a series of thermal resistors comprised of the resistances at interfaces and the intrinsic resistance of the materials, similar to the schematic shown in Fig. \ref{fig:Fig_01} (c). The thermal resistance we measure, due to the thickness of the film stacks, is the total resistance between the Ru and Si. Using different thicknesses of \textit{a}-GST, which varies the relative contribution of each resistor to the overall measured conductance, allows us to assess the relative contribution of each thermal resistance to the overall device thermal transport.

Figure \ref{fig:Fig_01} (b) shows the thermal conductance between Ru and Si, including all intermediate layers (spacer/ \textit{a}-GST/spacer), as a function of \textit{a}-GST thickness. The spacers we utilize are W, with thickness of 2 and 5 nm, amorphous SiO$_2$ (5 nm), and amorphous SiN$_x$ (5 nm), where the spacers' thicknesses are identical on either side of the \textit{a}-GST. As the thickness of \textit{a}-GST increases, the effect of the spacers on the overall thermal transport becomes negligible owing to the fact that the \textit{a}-GST layer becomes the dominant resistor. Based on Fig. \ref{fig:Fig_01} (b), for thicknesses greater than $\sim$10 nm, thermal conductance is largely governed by the \textit{a}-GST layer regardless of the adjacent material, whereas for thicknesses less than 10 nm, the effect of thermal boundary conductance become appreciable. Note, for SiN$_x$ and SiO$_2$ spacers, their thermal conductances are similar and lower than that of W. This is expected as the thermal conductivities of SiN$_x$ and SiO$_2$ are similar and more than an order of magnitude lower than that of W \cite{gaskins2017investigation}. However, it is important to note that the thermal conductance of the stack with the 5 nm W spacer is greater than that with 2 nm spacer. This is contrary to expectations when considering diffusive thermal transport processes, where thermal conductance decreases linearly with an increase in  thickness of the material. The observed reduction in thermal conductance for 2 nm W is attributed to the scattering of electrons and phonons at its boundaries. Similar thermal boundary conductance dependencies on the thickness of intermediate layer have been observed across Au/Ti/sapphire \cite{olson2018influence}, Au/Cr/sapphire \cite{jeong2016enhancement}, and Au/Cu/sapphire \cite{jeong2016enhancement} interfaces. The thermal conductivity of \textit{a}-GST is determined from these measurements by fitting a linear regression to the slope of the measured thermal resistance as a function of thickness, depicted in the inset of Fig. 1 (b). The thermal conductivity of \textit{a}-GST is determined to be 0.15 $\pm$ 0.02 W m$^{-1}$ K$^{-1}$, in good agreement with previously reported values \cite{lyeo2006thermal, risk2009thermal, lee2011thermal}.

\begin{figure*}
\includegraphics[scale = 0.5]{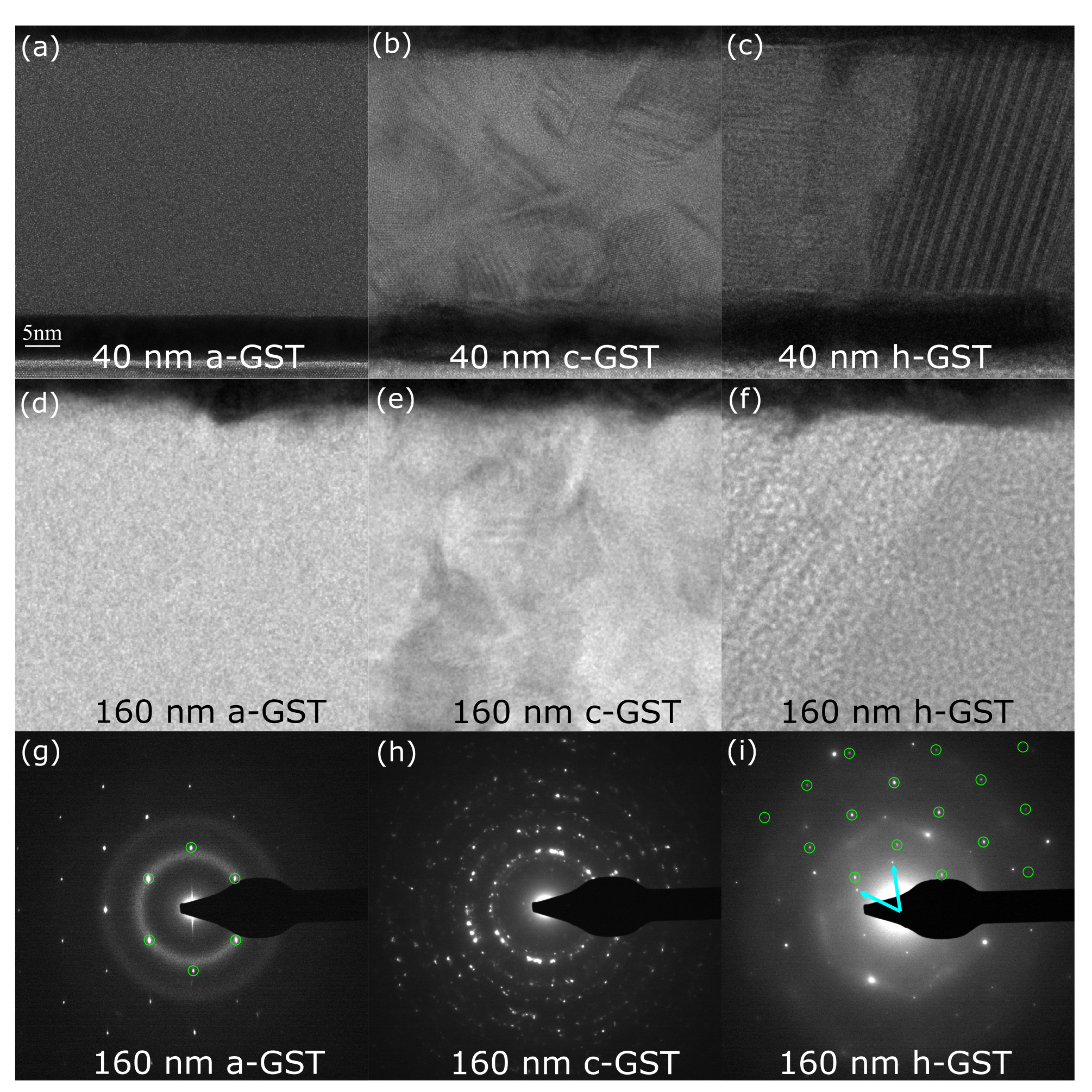}
\centering
\caption{Bright-field images of the: (a-c) 40 nm and (d-f) 160 nm GST thin films at: (a,d) 25 \degree C, (b,e) 240 \degree C and (c,f) 400 \degree C, showing a sequence of phase transformations from amorphous to cubic to hexagonal GST, respectively. Selected area diffraction patterns (SADP) of the 160 nm film at: (g) 25, (h) 240, and (i) 400°C, reflecting the phase transformations and microstructures observed in the BF images. In the inner ring of (g) and the top half of (i), the green circles indicate Bragg spots in the [110] zone axis of the Si substrate, and blue arrows in (i) show \{100\} primitive reciprocal-lattice vectors of h-GST in a [001] zone axis. Diffuse streaks in (i) extending through the \{2$\overline{1}$0\} Bragg spots are due to the transnational shear faults seen in (c) and (f).
\label{fig:Fig_02}}
\end{figure*}

\textbf{GST morphology at different phases.}
In order to confirm phase transformation and the quality of the crystal structure associated with each phase, we perform TEM on the 40 and 160 nm GST with \textit{in situ} heating (Figs. 2 (a)-(f)). The transition from diffuse rings in the selected area diffraction (SADP) to sharp diffraction rings denotes the transformation from \textit{a}-GST to polycrystalline cubic GST (c-GST), as shown in Figs. \ref{fig:Fig_02} g and h, respectively. This is in agreement with previous results showing that GST transforms from an amorphous phase to a face-centered cubic (FCC) lattice at $\sim$155 \degree C \cite{koenig2007thermoelectric, luo2019x, adnane2017high}. The 160 nm GST film thickness was measured as 160, 152, and 149 nm at 25, 240, and 400 \degree C, respectively, and similarly the 40 nm GST film thickness as 38.7, 36.9, and 33.7 nm at 25, 200, and 400 \degree C. On average, the thickness of the GST film changes by $\sim$5$\%$ and $\sim$6$\%$, at the transition from amorphous to cubic and cubic to hexagonal, respectively, which are comparable to the values of 6.5$\%$ and 8.2$\%$ reported elsewhere \cite{njoroge2002density}. The grain size for c-GST ranges from 10-20 nm in both films. When the films transform to the hexagonal phase, there are highly faulted grains that span the thickness of the films, as shown in Fig. \ref{fig:Fig_02} (c). A few differences in the lateral (i.e., in-plane) grain size were observed between the 40 and 160 nm films. In the 160 nm film, the lateral size of the smaller grains is approximately 50 nm, whereas in the 40 nm film, the size varied from 50 to 100 nm. Large faulted grains in the 160 nm film are $\sim$100-200 nm wide, but only about 100 nm wide in the 40 nm film. The cross-plane dimension of the grains in the 40 nm film are often that of the film thickness (i.e., 40 nm), while there is a range in the 160 nm film. The SADP from GST at 400 \degree C (Fig. \ref{fig:Fig_02} (i)) displays a single GST [001] zone axis and the [110] zone axis of the Si substrate, due to the much larger grain size compared to 240 \degree C (Fig. \ref{fig:Fig_02} (h)). Diffuse streaks emanate from the \{2$\overline{1}$0\} Bragg spots as a result of the faulted grain. Highly faulted structures have been observed in a variety of chalcogenides including Ge-Sb-Te compounds \cite{da2008insights, wang2018genesis, welnic2007phase, lotnyk2017atomic, mio2019impact, behrens2018ultrafast}. The large, faulted grains grew laterally by growth ledges that often nucleate at the W/GST interface. The growth ledges then propagate along the h-GST/c-GST interface, consuming smaller grains as they move. Crystalline GST is composed of van der Waals coupled building blocks, each of which contains five Te layers separated by either a Ge or Sb anion layer \cite{da2008insights, wang2018genesis, wang20182d, wang2018unconventional}. By shifting each nine-layer building block by a partial lattice vector, the c-GST becomes h-GST and vice versa. Recent literature suggests that the weak bonding between the sesqui-chalcogenides building blocks, such as Sb$_2$Te$_3$, significantly exceeds those of van der Waals forces and, therefore, possess more of a metavalent bond nature \cite{cheng2019understanding, mio2019impact}. As a result of the weak bonding between the blocks, there is a low energy barrier to passing partial dislocations that transform the lattice and cause faults. Faulted grains in the 160 nm sample were primarily at an angle to the film normal as seen in Fig \ref{fig:Fig_02} (f). The same was observed for the 40 nm film, such as on the right side of Fig \ref{fig:Fig_02} (c) in addition to grains whose building blocks/faults were parallel to the film normal, as on the left side of Fig. \ref{fig:Fig_02} (c).  

\textbf{Elevated temperature thermal properties of GST.}
Above, we showed that the effects of room temperature TBC values on overall device thermal resistance are only appreciable for \textit{a}-GST when the thickness is less than 10 nm. However, as the \textit{a}-GST film changes phase, its intrinsic thermal conductivity increases by almost an order of magnitude. This implies that thermal transport in the crystalline phase should be more dramatically affected by the TBC than in the amorphous phase. Therefore, it is crucial to understand the effects of thermal transport across W/GST interfaces as GST undergoes phase transitions. TDTR measurements are taken as a function of temperature using a resistive heating stage that allows us to measure the thermal conductivity of GST and the thermal boundary conductance at the h-GST/W interface from room temperature up to 400 \degree C, thereby, capturing the thermal properties of GST in all of phases (i.e., amorphous, cubic and hexagonal). Although cross plane electrical resistivity measurements for GST films are beyond the scope of this paper, in supplementary note 2 we use available electrical resistivity data in the literature to differentiate the contributions of electrons and phonons to the total thermal conductivity of GST.

\begin{figure*}
\includegraphics[width=\textwidth]{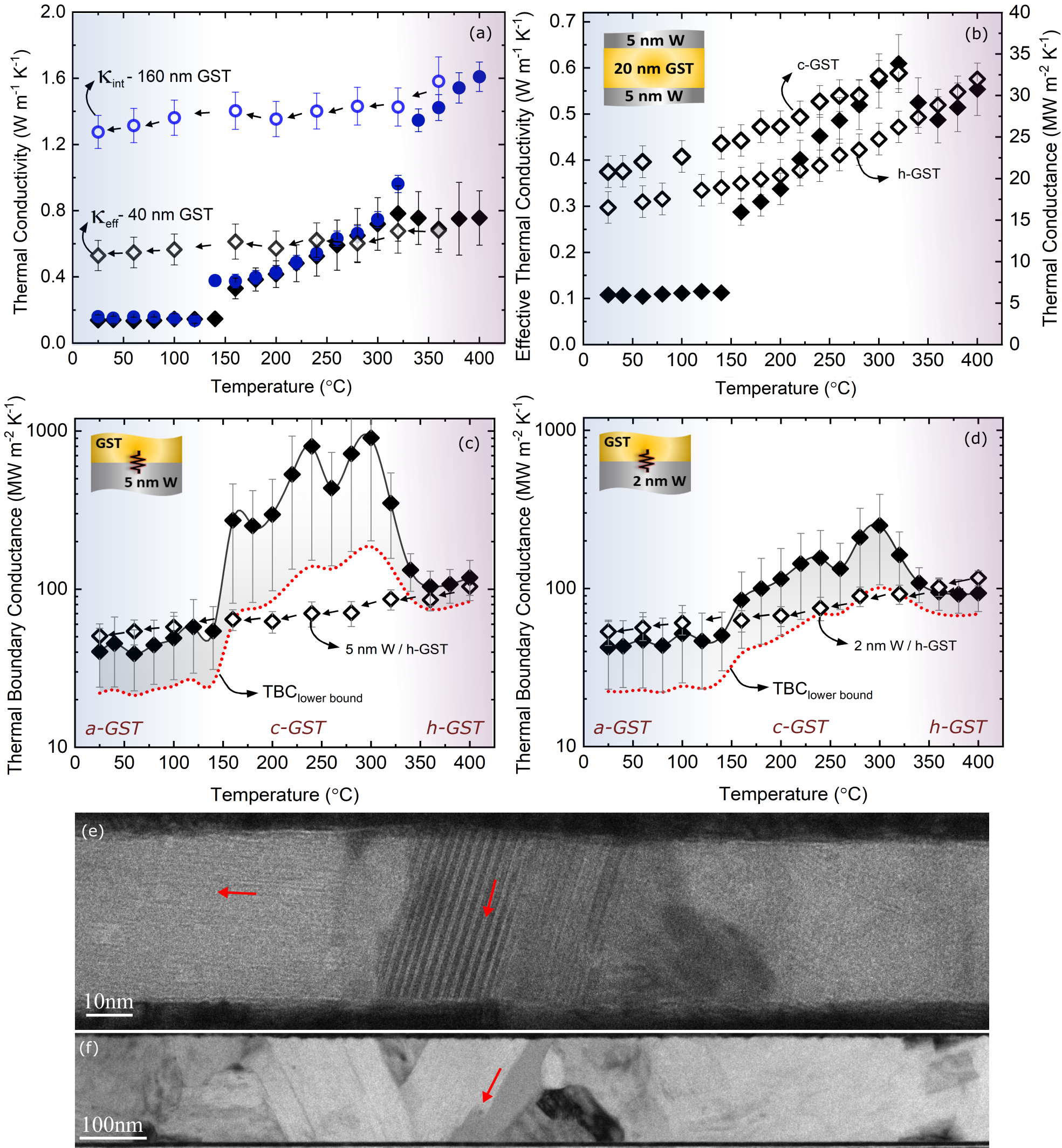}
\centering
\caption{(a) Thermal conductivity of GST layer sandwiched between 5 nm W spacers for 40 nm (diamonds) and 160 nm (circles) GST films. The solid symbols correspond to the thermal conductivity of GST as it transitions through different phases upon heating and hollow symbols correspond to the thermal conductivity of h-GST upon cooling. (b) Total thermal conductance across Ru/5nm W/20 nm GST/5 nm W/Si as a function of temperature for different phases of GST upon heating and cooling when annealed to 300 and 400 \degree C. (c,d) Thermal boundary conductance for as-deposited GST upon heating and h-GST upon cooling with 5 and 2 nm W spacer, respectively. The discussion regarding uncertainty calculation is given in supplementary note 1. (e,f) Bright field images of 40 nm and 160 nm GST films at 400 \degree C.
\label{fig:Fig_03}}
\end{figure*}

Figure \ref{fig:Fig_03} (a) shows the thermal conductivity of 40 nm and 160 nm thick \textit{a}-GST layers that are heated under nominally identical conditions across various temperatures. In this figure, the solid symbols correspond to the thermal conductivity of \textit{a}-GST when heated from room temperature up to 400 \degree C, whereas the hollow symbols correspond to the thermal conductivity of h-GST when cooled down from 400 \degree C to room temperature. The solid circles denoting the 160 nm film in Fig. \ref{fig:Fig_03} (a) show a clear transition from \textit{a}-GST to c-GST, and c-GST to h-GST at approximately 150 \degree C and 340 \degree C, respectively, in good agreement with reported literature values \cite{lyeo2006thermal, reifenberg2007thickness, lee2013phonon}. The enhancement of thermal conductivity in the crystalline phase is attributed to the dissolution of disordered vacancy clusters and increasing order in the crystalline phase \cite{siegrist2011disorder, zhang2012role}. After the sample reaches 400 \degree C and the GST is fully transformed into the hexagonal phase, its thermal conductivity is measured as the sample is cooled down to room temperature, shown as hollow circles in Fig. \ref{fig:Fig_03} (a). The thermal conductivity of the 160 nm h-GST decreases slightly over temperature as a result of reduced electronic contribution to the thermal conductivity \cite{siegrist2011disorder}. 

On the other hand, for the 40 nm thick GST, the measurement of intrinsic thermal conductivity in the crystalline phase is increasingly difficult as the effects of TBC dictate different measurement sensitivities to thermal conductivity as compared to the 160 nm case (see supplementary note 1). For this reason, we report the effective thermal conductivity (k$_{eff}$ = G$_{Ru/W/GST/W/Si}$ $\times$  d$_{GST}$), depicted as solid diamonds, which incorporates both the effects of the intrinsic thermal conductivity of GST and the associated TBCs. The effective thermal conductivity for the 40 nm thick GST sample follows a similar trend to that of 160 nm film up to 300 \degree C, except for the slight upward shift in crystallization temperature to 150 \degree C. The agreement of thermal conductivity between the two thicknesses is due to lower sensitivity of our measurement to TBC in the amorphous and cubic phase. However, upon transformation from c-GST to h-GST, the TBC at the h-GST/W interface considerably decreases. As a result, we observe that the effective thermal conductivity for the 40 nm GST film deviates from the 160 nm GST above 300 \degree C. Above 300 \degree C we no longer measure the intrinsic thermal conductivity of the h-GST layer but, instead, a convolution of the h-GST thermal conductivity and the h-GST/W TBC. The effective thermal conductivity for the 40 nm sample plateaus near $\sim$0.8 W m$^{-1}$ K$^{-1}$, almost a factor of two lower than the thermal conductivity measured for 160 nm h-GST. This difference is even more pronounced when the samples are cooled down to room temperature where the thermal conductivity for 40 and 160 nm is $\sim$0.5 and $\sim$1.25 W m$^{-1}$ K$^{-1}$. In order to ensure the observed reduction in the effective thermal conductivity is not due to any microstructural changes in the film, we present extensive TEM with \textit{in situ} heating to compare the quality of the crystals for both thicknesses. Although defects, such as stacking faults, occur in the hexagonal phase as is shown in Fig. \ref{fig:Fig_03} (e,f), we did not identify any significant microstructural anomalies between the two cases that may explain such a significant reduction in the 40 nm thick GST. The TEM results imply that the intrinsic thermal conductivity in both cases remains unaltered and, therefore, the observed discrepancy must be related to extrinsic effects such as TBC. This finding indicates that, contrary to what is generally assumed, total thermal transport does not necessarily increase with the increase in thermal conductivity of GST.

To further support our hypothesis regarding the effect of TBC on thermal transport in h-GST, we measure the total thermal conductance across the Ru/W/GST/W/Si film stack for a 20 nm thick GST layer. For this thickness regime, the Kapitza length, defined as the thermal conductivity divided by the TBC and represents the thickness of a material in which thermal boundary conductances can influence the overall thermal transport of a system, is comparable to the thickness of the film and, as a result, the effect of interfaces in our measurements are more pronounced compared to the 40 nm film. For this thickness, the thermal conductances as a function of temperature are depicted in Fig. \ref{fig:Fig_03} (b) in solid diamonds, which follow the same trend observed in effective thermal conductivity of 40 nm film with a more pronounced drop at the transition from c-GST to h-GST. This is clear evidence for the opposite trend of TBC to that of the thermal conductivity for c-GST to h-GST transition. In order to compare the TBC for c-GST vs. h-GST, we take another 20 nm thick \textit{a}-GST sample and heat it up to 320 \degree C where the GST film becomes fully cubic. By cooling the sample down to room temperature, we measure the thermal conductance in the cubic phase as a function of temperature (hollow squares). Once the thermal conductance over the temperature range of interest was measured, the same sample is again heated up to 400 \degree C to transform the c-GST into hexagonal phase and its thermal conductance was remeasured upon cooling (hollow diamonds). As shown in Fig. \ref{fig:Fig_03} (b), we obtain a higher thermal conductance in c-GST than that of the h-GST phase across the entire temperature range. This is contrary to expectations as the thermal conductivity in h-GST is nearly two times higher than the c-GST, however, due to relatively poor thermal transport at the interfaces, a lower thermal conductance is measured. Based on the results presented here, we conclude that the TBC between h-GST and W is lower than that of the c-GST and W.

\begin{figure*}
\includegraphics[width=\textwidth]{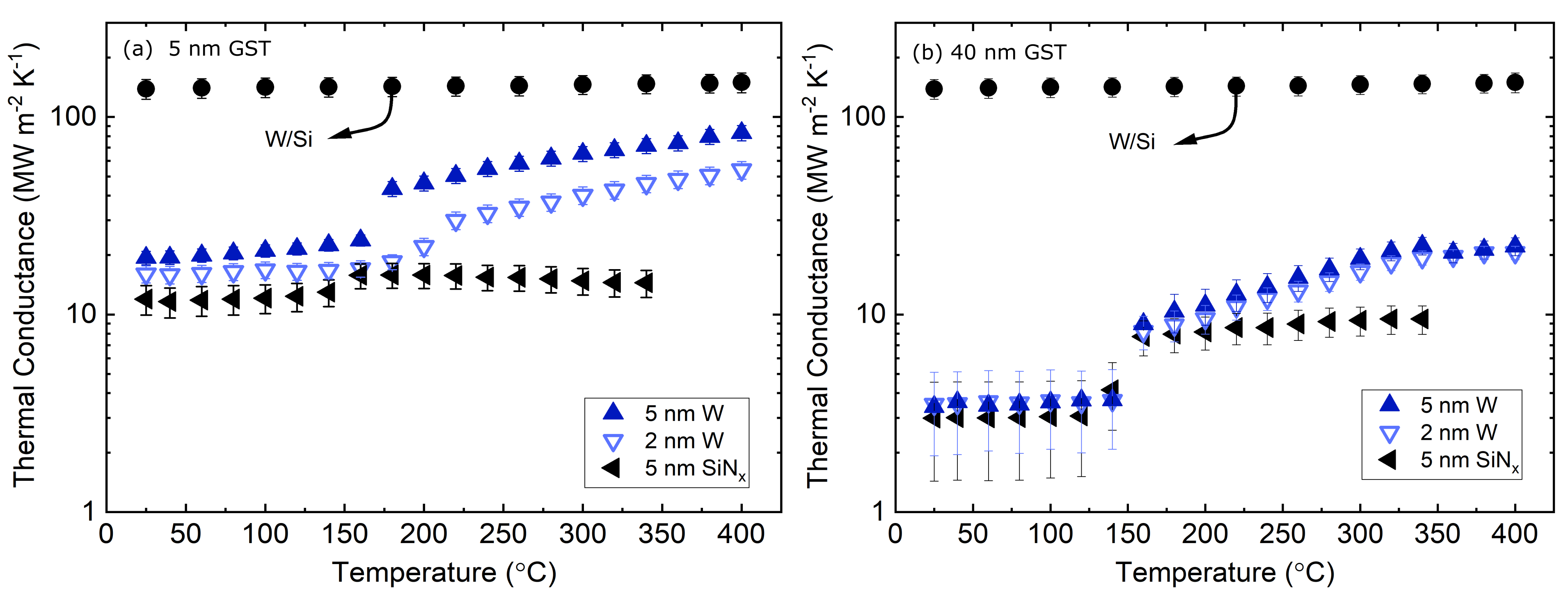}
\centering
\caption{Thermal conductance for (a) 5 nm and (b) 40 nm thick GST in contact with different spacers. Experimental data are not available for SiN$_x$ above 340 \degree C due to film delamination.
\label{fig:Fig_04}}
\end{figure*}

In this respect, due to the significant impact of TBC on thermal transport of thin film GST, it is important to study how it changes across various phases. Figure \ref{fig:Fig_03} (c,d) shows the TBC between GST and two different thicknesses of W. The TBC in h-GST is significantly suppressed compared to c-GST. A similar reduction in TBC has previously been observed between GST and aluminum films across the cubic to hexagonal phase transition which was attributed to the inter-diffusion of GST constituents into the aluminum layer and formation of 2 nm interfacial layer \cite{battaglia2013thermal}. However, according to our TEM images, we do not observe an additional interfacial layer after the c-GST to h-GST phase transition. Further, it has been shown that the degree of disorder in GST decreases as the annealing temperature increases and GST transitions into hexagonal phase \cite{siegrist2011disorder,zhang2012role}. With that in mind, we use a simplistic model via molecular dynamics simulations and demonstrate that a change in atomic-scale disorder at the interface from c-GST to h-GST can, in fact, be another possible reason behind the suppression of thermal transport.  Disorder and defects at interfaces are well known to influence the TBC, and have in fact been computationally and experimentally shown to enhance TBC \cite{tian2012enhancing, english2012enhancing, gorham2014ion, giri2020review}. To this end, our molecular dynamics simulations suggest that varying atomic scale interfacial disorder could explain the TBC change across the c-GST to h-GST transition, and this disorder plays a stronger role in the changes observed in TBC than the transition in the crystal lattice and phonon density of states. We note that in our molecular dynamics simulations, we are using Lennard-Jones potential that are not developed to predict the thermal properties of W or GST. However, the simplicity of these potentials allows us to assess our hypotheses to general classes of materials, thus providing means to broadly study our posits of the origin of reduction in TBC across the cubic to hexagonal transition (see supplementary note 3). In Fig. \ref{fig:Fig_03} (c,d) the dotted line represents the minimum limit to the TBC by assuming the worst case scenario for the effective parameters based on 10$\%$ uncertainty, which in many cases is far higher than measured uncertainty. Lack of sensitivity in the amorphous and cubic phases causes the reported range of TBC (best-fit to minimum-limit) to be quite broad, but as the GST transitions to hexagonal and gains more sensitivity to TBC, this range contracts. The hollow diamonds in Fig. \ref{fig:Fig_03} (c,d) shows the TBC for h-GST which decreases by almost a factor of two as the sample is cooled down from 400 \degree C to room temperature. A step by step description for our approach to calculate the TBC is given in the supplementary note 1. Figures \ref{fig:Fig_03} (c,d) again demonstrate that the TBC for 5 nm W is higher than that of 2 nm W, especially in the crystalline phase where the effect of TBC is more pronounced. This difference becomes even more appreciable for 5 nm GST at elevated temperature (see Fig. \ref{fig:Fig_04}(a)). We note that a similar increase in TBC by increasing the thickness of an ultra-thin intermediate layer at a metal/non-metal interface has recently been observed \cite{jeong2016enhancement, olson2018influence, blank2018influence}.

Figure \ref{fig:Fig_04}(a, b) shows the thermal conductance across Ru/W/GST/W/Si as a function of temperature for 5 nm and 40 nm thick GST film with different spacers (2 nm W, 5 nm W, and 5 nm SiN$_x$). The thermal conductance for a 10 nm W control is also plotted to clarify that the TBC at Ru/W and W/Si interfaces are relatively constant and sufficiently large compared to GST intrinsic thermal conductivity and GST/W interface. In Fig. \ref{fig:Fig_04}(a), we observe a linear trend for thermal conductance of 5 nm GST film as a function of temperature after the crystallization onset ($>150$ \degree C) for W spacer. This is in contrast with the trend observed in Fig. \ref{fig:Fig_04}(b) for 40 nm thick GST where the thermal conductance plateaus above 300 \degree C. In a fully diffusive thermal transport regime, the effect of reduced TBC in h-GST must be even more noticeable for 5 nm GST where the effect of intrinsic thermal conductivity is minimum. To explain this, it has been shown that as the thickness of the GST layer decreases to ultra-thin, the onset of crystallization increases to higher temperatures \cite{chong2007thickness}. As a result, it is tempting to attribute this increase to a crystallization lag where at 300-400 \degree C range the film is gradually transitioning to h-GST. To assess this hypothesis, we can predict the total thermal conductance of the Ru/ 5 nm W/5 nm GST/ 5 nm W/Si stack at different phases using thermal conductivity and TBC measured in the previous section. Using a series resistors model, we calculate the stack total thermal conductance at 400 \degree C to be 43 $\pm$ 5 MW m$^{-2}$ K$^{-1}$ which is almost a factor 2 lower than the measured value of 83 $\pm$ 7 MW m$^{-2}$ K$^{-1}$. It is noteworthy to mention that the TBC for GST/W and W/GST interfaces alone, is 59 $\pm$ 7 MW m$^{-2}$ K$^{-1}$, which is significantly lower than the measured thermal conductance for the entire stack. The fact that we measure almost a factor of two higher thermal conductance for h-GST at 400 \degree C cannot be explained within the diffusive thermal transport limit. On the other hand, it has been shown that in bulk tungsten the average electron mean free paths before scattering with phonons at room temperature can be as long as 19.1 nm \cite{choi2012electron}. Additionally, due to tungsten high lattice thermal conductivity ($\sim$46 W m$^{-1}$ K$^{-1}$)\cite{choi2012electron}, it has phonons with long mean free paths relative to other metals \cite{chen2019understanding}. From this, we conservatively estimate the phonon mean free path in tungsten to be on the order of $\lambda$ = 3k$_p$/Cv = (3 $\times$ 46)/(2.58$\times$10$^6$ $\times$ 5174) = 10.3 nm. Since the thickness of 5 nm GST is within the range of the phonons' and electrons' mean free paths, we attribute this enhancement in thermal conductance to ballistic transport of energy carriers emitted from the top W spacer to the bottom W spacer.

We further study this hypothesis by measuring the thermal conductance of a similar multilayer system in which we replace the W spacer with amorphous SiN$_x$. The SiN$_x$ spacer is widely used as a dielectric in electronic devices due to its high electrical and thermal resistivity \cite{gaskins2017investigation}. As a result, we expect the contribution of phonons and electrons to thermal conductance in SiN$_x$ to be negligible compared to that of W spacer. As shown in Fig. \ref{fig:Fig_04}(a) the thermal conductance across the Ru/SiN$_x$/5 nm GST/SiN$_x$/Si stack is relatively constant across all temperatures, which gives further credence to our hypothesis regarding ballistic electron/phonon transport leading to increased thermal conductance in the 5 nm GST films between W contacts. In this case, no significant enhancement is observed in thermal conductance across Ru/Si after the crystallization temperature. This suggests that, due to the absence of long-wavelength electrons and phonons in amorphous SiN$_x$, there is no ballistic transportation from the top SiN$_x$ to the bottom SiN$_x$ layer.

\begin{figure}
\includegraphics[width=\columnwidth]{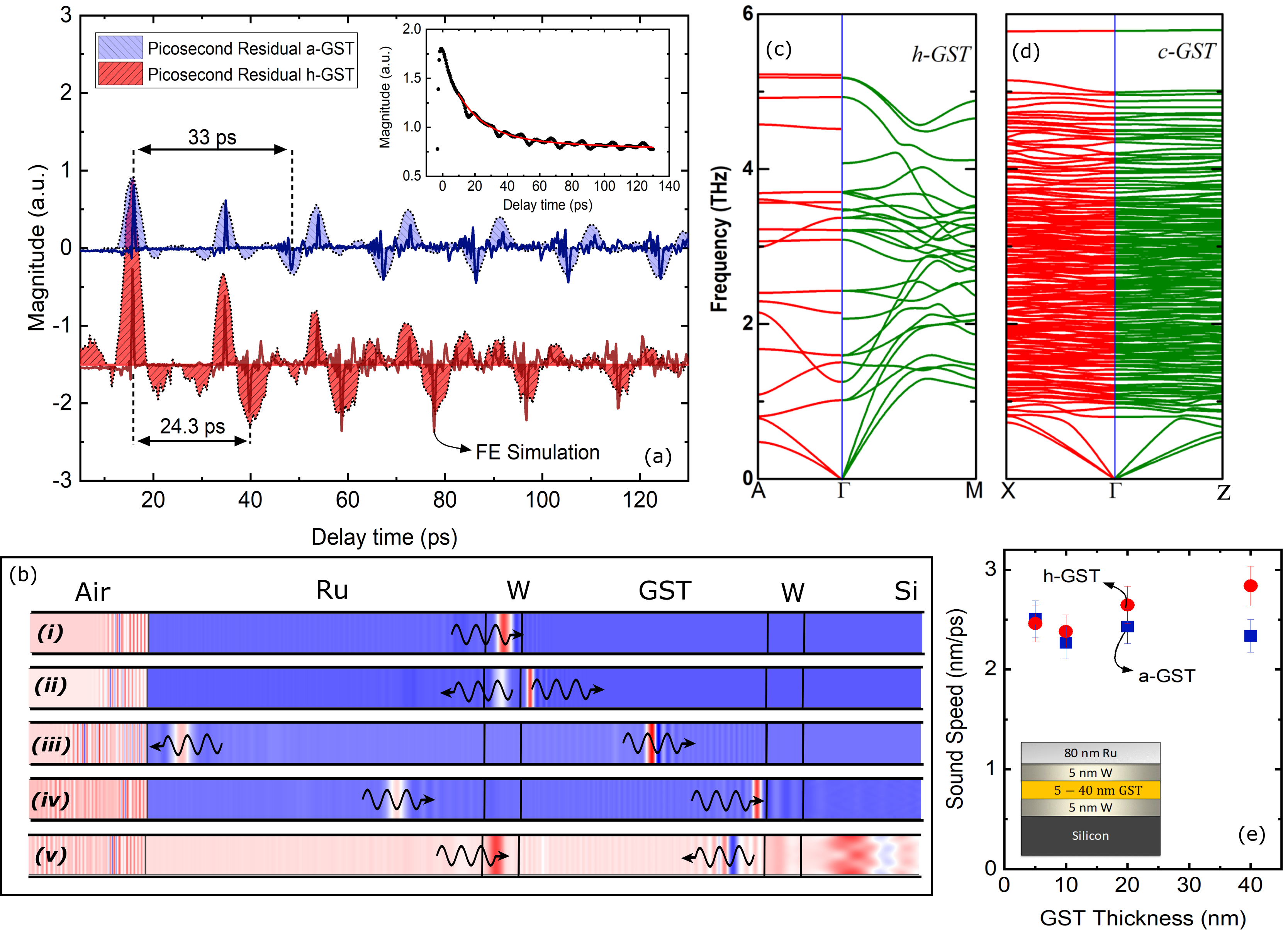}
\caption{(a) Picosecond ultrasonic measurements for 40 nm of \textit{a}-GST (blue) and 400 \degree C annealed h-GST (red). The “humps” and “troughs” corresponds to the reflection of strain waves off of W/GST and GST/W interfaces, respectively. The lines correspond to finite element simulation of strain wave propagation across different layers. (b) A representation of strain wave and how it propagates and reflects off of various interfaces. (c,d) Calculated phonon dispersions for h-GST and c-GST. (e) Sound speed measurements for different thicknesses of \textit{a}-GST and h-GST.
\label{fig:Fig_05}}
\end{figure}

\textbf{Sound speed measurements.} In order to gain further insight into the thermal properties of GST layers below 40 nm, picosecond ultrasonic measurements are used to measure the sound speed of the GST thin films. In this measurement, the absorption of the ultrashort laser pulse launches a strain wave from the sample surface due to the rapid heating. This results in qualitative “humps” and “troughs” superimposed on the TDTR thermal decay curve. The temporal spacing of these picosecond ultrasonic signals is related to the time it takes for the strain waves to travel through a material and reflect off of sub-surface interfaces as demonstrated in schematic in \ref{fig:Fig_05} (b). Thus, with knowledge of the thicknesses of each film (determined via TEM), we can estimate the sound speed. For a better interpretation of picosecond ultrasonic data (the inset in Fig. \ref{fig:Fig_05} (a)), the thermal decay curve is subtracted from the best exponential fit to the experimental data and the residual is presented in Fig. \ref{fig:Fig_05} (a), where the “humps” and “troughs” in the plot are the consequence of strain waves reflection from the top and bottom interfaces of the GST layer.

By measuring the time between the upward “humps” and downward “troughs” in Fig. \ref{fig:Fig_05} (a), we obtain the time it takes for the strain waves to travel across the GST. Based on the measured thickness of the GST film, the longitudinal sound speed in a 40 nm \textit{a}-GST and h-GST layer is measured to be 2300 $\pm$ 200 m s$^{-1}$ and 2800 $\pm$ 200 m s$^{-1}$, respectively. Although, our sound speed measurement for \textit{a}-GST agrees with literature, the sound speed for h-GST is below what has been reported \cite{lyeo2006thermal, mukhopadhyay2016optic}. To further understand the discrepancy between our sound speed measurements and previously reported value in hexagonal phase, we perform first-principle calculations for GST in both cubic and hexagonal phase. Figure \ref{fig:Fig_05} (c) and (d) show the phonon dispersion for h-GST and c-GST. For the cubic phase, the average group velocities for the LA and TA modes were calculated as 3531 m s$^{-1}$ and 1658 m s$^{-1}$, respectively. Due to the structural anisotropy, we observe strong anisotropic phonon dispersion in h-GST, giving anisotropic group velocities along the in-plane (LA= 3828 m s$^{-1}$; TA = 2398 m s$^{-1}$) and out-of-plane directions (LA= 3502 m s$^{-1}$; TA= 2567 m s$^{-1}$). The similar sound speed between c-GST and h-GST along the out-of-plane direction is consistent with our observation from the picosecond ultrasonic measurements. However, quantitatively, higher LA group velocity in our calculation compared to the measured value is in contrast with the typical under-binding tendency of the generalized gradient approximation that increases bond length and softens the phonons leading to lower group velocities compared to that in measurements. This implies that the LA mode velocities in h-GST and c-GST thin films are lower than that in their bulk counterpart. This reduction is more prominent when the sound speed for different thicknesses are measured and plotted in Fig. \ref{fig:Fig_05} (e). As can be seen, as the thickness of GST decreases the sound speed in hexagonal phase converges to that of the amorphous phase. The reduced measured sound speed for h-GST is most likely due to the existence of amorphous regions near the interface, and we hypothesize not intrinsic to the GST. For more details on sound speed measurements refer to the Supplementary Note 4.

\textbf{Discussion.} As the memory cell dimension in PCM devices shrinks and progress towards superlattice structures, it is essential to account for the parameters that are not conventionally considered in thick regime such as interfacial thermal resistance and ballistic thermal transport. In superlattice PCMs, due to existence of several interfaces in a single cell, engineering the interfacial resistance can substantially improve the performance of the device. For superlattice structures, the TBR at GeTe/Sb$_2$Te$_3$ interface is reported \cite{okabe2019understanding}  to be around 3.4 m$^{2}$ K GW$^{-1}$. Our work, in addition to reporting a significantly higher TBR between h-GST and W, $\sim$10 m$^{2}$ K GW$^{-1}$, demonstrates how judiciously engineering the interface between GST and its adjacent material can reduce the reset current. To demonstrate the effect of TBR on thermal transport, we show that for a 20 nm thick GST film the effective thermal conductivity can be reduced by a factor of 4 due to the increased interfacial resistance in h-GST. Our work shows that interfacial resistance is only effective in reducing thermal transport when the GST thicknesses is less than 40 nm. On the other hand, there is a limitation on reducing the thickness of the GST layer before the thermal transport transitions into a ballistic regime. According to our results, as the thickness of GST reaches $\sim$5 nm, ballistic transport of phonons/electrons from the top W electrode to the bottom electrode increases the thermal transport by almost a factor of two. To prevent this ballistic transport effect, it is important to choose interlayers that have carriers with short mean free paths. In our previous work \cite{aryana2020thermal} we demonstrated that materials such as CN$_x$ with short mean free path energy carriers can serve as a better electrode than tungsten when the device dimension reaches below 10 nm. For the specific layer configuration studied here, W/GST/W, the GST thickness at which electrode engineering has the biggest impact in efficiency optimization is approximately 20 nm. For thinner GST thicknesses, ballistic thermal transport limits thermal confinement and at larger thicknesses, bulk properties of the GST will play a larger role and as a result the effect of TBR between GST and the electrode diminishes. 

\begin{figure*}
\includegraphics[scale = 0.9]{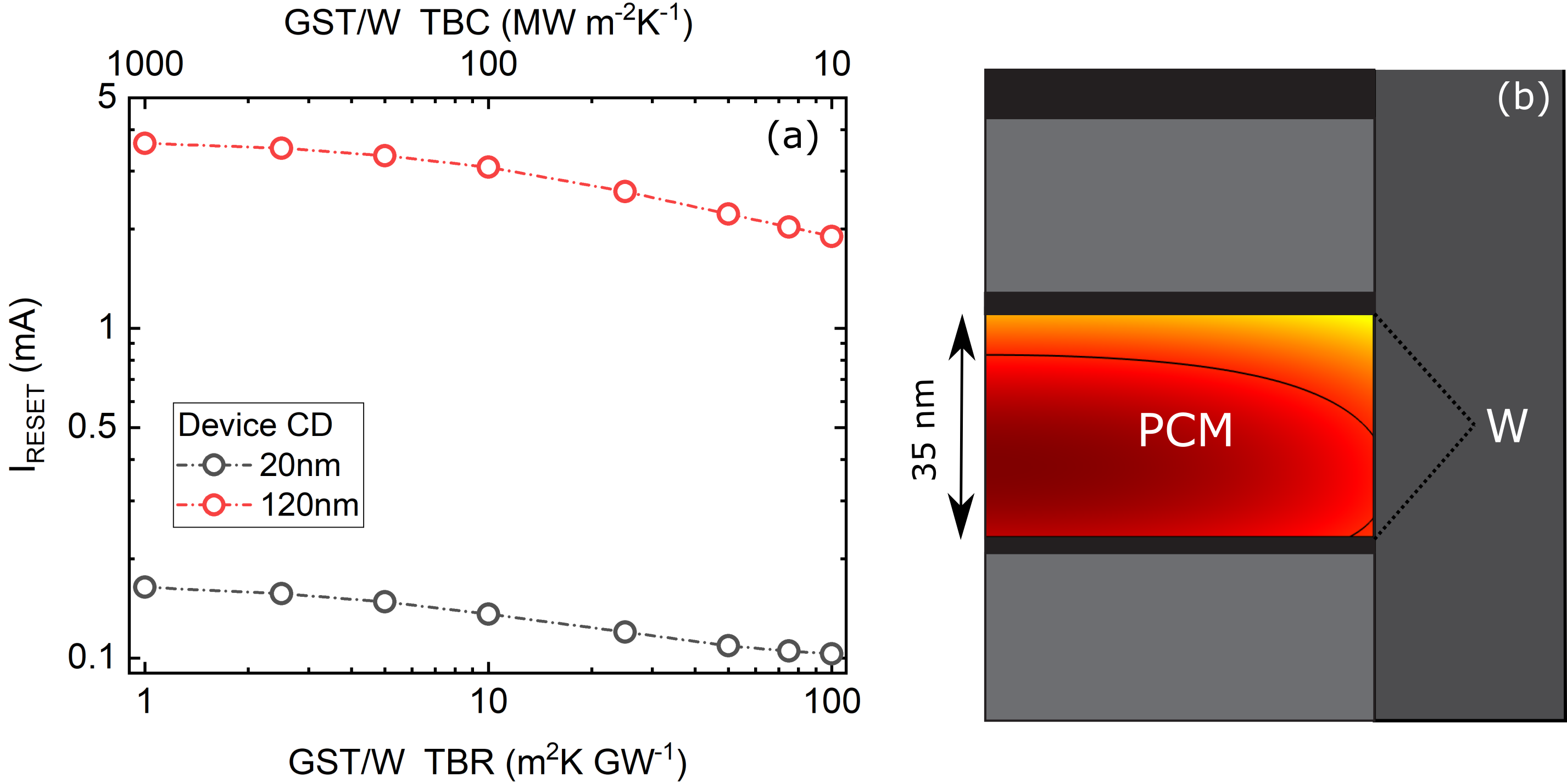}
\centering
\caption{ (a) Simulation results for the reset current as a function of thermal boundary resistance (TBR) between GST and W for two different device lateral sizes, (b) Schematic for the PCM configuration and its corresponding temperature gradient.
\label{fig:Fig_06}}
\end{figure*}

Earlier, we demonstrated that by reducing the W thickness from 5 to 2 nm, thermal conductance can be moderately suppressed. In order to demonstrate the effect of W layer thickness on PCM device performance, we use computational models for a PCM device in confined cell geometry. For this, a 35 nm thick GST unit is sandwiched between identical W layers (2 or 5 nm), and connected to TaN electrodes. The cell geometry is a cylinder confined by dielectric materials; we repeat our simulation for two different cell dimensions with lateral size of 20 nm and 120 nm diameter in order to compare thermal transport in small and large devices. The simulations are carried out by using finite-element simulation package COMSOL Multiphysics. Table 1 summarizes the step-by-step simulation process as we progressively add measured parameters into the simulation. Our simulations demonstrate that thinning the W layer from 5 nm to 2 nm, taking 35$\%$ reduction in thermal conductance into account, leads to reset current (I$_{reset}$) reduction from 133 $\mu$A to 127 $\mu$A for the 20 nm device and from 3.07 mA to 2.88 mA for the 120 nm device, corresponding to 4.5$\%$ and 6.2$\%$ reduction in reset current, respectively. Although manipulating W thickness leads to a modest reduction in the reset current, it should be noted that this is achieved through practical changes in an interface that is not typically optimized for its thermal properties. Further optimization along these lines could lead to larger improvements. In order to demonstrate this, we extend our simulations to account for a range of TBR between the phase change unit and the adjacent electrode. It is expected that the TBR between GST and most materials to fall in the range of 1-100 m$^2$ K GW$^{-1}$ \cite{reifenberg2008impact}. The result of our simulations for reset current as a function TBR between PCM/electrode and the cell configuration for a 120 nm confined cell is presented in Fig. \ref{fig:Fig_06} (a) and (b), respectively. Our predictions suggest that the reset current can be reduced up to $\sim$40\% and $\sim$50\% depending on the device lateral size if the TBR changes from 1 to 100 m$^2$ K GW$^{-1}$. In superlattice structures where there are multiple interfaces the reset current can be even further reduced. Boniardi \textit{et al.} \cite{boniardi2019evidence} observed nearly 60\% reduction in set and reset current for (GeTe/Sb$_2$Te$_3$)/Sb$_2$Te$_3$ superlattice compared to bulk GST, which they attributed to increased thermal resistances in the superlattices from the period interfaces as compared to the GST. Our results highlight the importance of interfacial engineering on thermal confinement of PCM memory cells.

In summary, we reported on the thermal properties of GST for thicknesses below 40 nm and compared the results against the thick film regime (160 nm). We demonstrated that as the length scale of phase change memory cells decrease to the dimensions of the order of carriers’ mean free paths, the mechanism of heat transport drastically differs from its bulk. In addition, our results demonstrate that as the GST transition from one crystallographic phase to another, the interfacial resistance changes. The TBR for \textit{a}-GST/W, c-GST/W, and h-GST/W interfaces are measured to be approximately 25 $\pm$ 5, 3 $\pm$ 1.5, 10 $\pm$ 2 m$^2$ K GW$^{-1}$, respectively. Our molecular dynamics simulations results suggest that a change in phase from cubic to hexagonal does not significantly change the thermal boundary conductance. However, structural disorder at the interface plays an important role in the reduction of TBC from the cubic to the hexagonal phase. Overall, the interfacial resistance for a 20 nm thick GST film results in a factor of 4 reduction in the effective thermal conductivity from $\sim$1.25 to $\sim$0.3 W m$^{-1}$ K$^{-1}$ at room temperature. Our work illustrated that the thermal boundary resistance can be employed to substantially suppress heat transport in phase change units. We use simulations to elucidate the effect of TBR on the reset current for two different cell sizes. According to these simulations, the TBR can lead up to 40\% and 50\% reduction in reset current. The results presented in this work improve our knowledge of thermal transport mechanism in ultra-thin phase change units and enable us to design PCM devices with superior performance.

\begin{table}
  \caption{The impact of parameters extracted from the empirical measurements on the modeling results. The parameters in red are taken from the measurements in this study.}
  \label{tbl:excel-table}
  \includegraphics[width=\linewidth]{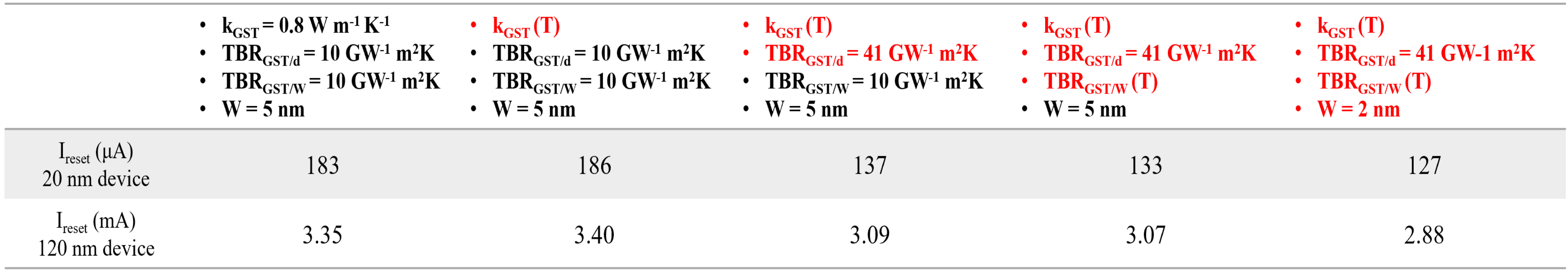}
\end{table}

\section*{Methods}
TDTR is a non-contact optical measurement technique that utilizes a modulated pump beam to create an oscillatory temperature rise at the surface of the sample, and a probe beam to measure the changes in the thermoreflectance of the surface due to the thermal excitation. In our two-tint TDTR configuration, the output of an 80 MHz Ti:sapphire femtosecond pulsed laser with a center wavelength of $808$  nm is split into two paths, a pump and a probe. The pump beam is passed through an electro-optical modulator (EOM) and modulated at a frequency of 8.4 MHz, while the probe beam is directed down a mechanical delay stage in order to induce a time-dependent signal relative to the arrival of pump pulses. The beams are combined by passing the probe through a high-pass filter and then both beams are directed through a 10X objective resulting in spot radius of 22 $\mu$m and 11 $\mu$m for the pump and probe at the sample surface, respectively. The uncertainty for room temperature measurements are determined by considering 7$\%$ change in the thickness of the Ru transducer. For calculating the uncertainty in thermal conductivity as a function of temperature for thin film samples, since the thickness of GST film alters across different temperature ranges, we assume 7$\%$ change in the thickness of GST instead of the transducer. 

For ab-initio density functional calculations, we considered two special quasirandom structures (SQSs) with 45 atoms (10 Ge, 10 Sb, and 25 Te atoms) to represent the c-GST and \textit{a}-GST. The \textit{a}-GST structure was obtained from a rapid quench after molecular dynamics simulations using GGA-PBE exchange-correlation functional form as implemented in VASP \cite{kresse1996efficiency, kresse1999ultrasoft}. The details of the structures can be found elsewhere \cite{mukhopadhyay2016competing}. For h-GST, we used the structure given by Kooi et al. \cite{kooi2002electron}  as it gives a more stable structure \cite{mukhopadhyay2016optic} compared to that given by Matsunaga et al. \cite{matsunaga2007structural}. The harmonic interatomic force constants (IFCs) were  calculated using VASP-phonopy \cite{togo2015first} interface with a 2×2×2 supercell where an energy cutoff of  500 eV was used. The calculated phonon dispersion for h-GST and c-GST is given in Fig. \ref{fig:Fig_05} (c,d). However, the calculated IFCs at 0K resulted in imaginary modes for \textit{a}-GST indicating that harmonic IFCs are not sufficient to describe the lattice dynamics of \textit{a}-GST.  

The finite element simulations of the strain wave propagation shown in Fig. \ref{fig:Fig_05} were implemented using Structural Mechanics and Acoustics module in COMSOL Multiphysics. To form a symmetric coherent wave, periodic boundary conditions were used for the top and bottom boundary along the Y-axis and low reflecting boundary condition on both ends along the X-axis. The material properties input for these simulations are density, Young's modulus, and Poisson's ratio. The density for the amorphous and crystalline state is assumed to be 5870 and 6200 kg m$^{-3}$. The Young's modulus is obtained from our empirical sound speed measurement where we calculate 32 GPa for \textit{a}-GST and 50 GPa for h-GST. The thickness of the layers are chosen to replicate the experimental values i.e. for Ru, W, GST, and Si the thicknesses are 80 nm, 5 nm, 40 nm, and semi-infinite, respectively. To create the strain waves, a short displacement pulse (half-sine) is applied to the surface of the Ru and the resulting echoes, generated due to the reflection of strain waves off of different interfaces are probed at the Ru surface in temporal resolution similar to picosecond ultrasonic measurement.

\textbf{Data availability.} The data that support the findings of this study
are available from the corresponding author on request.

\section*{Acknowledgements}
   We appreciate support from Western Digital Technologies, Inc. This manuscript is based upon work supported by the Air Force Office of Scientific Research under Award No. FA9550-18-1-0352. SM acknowledges support from NRC Research Associateship.

\bibliographystyle{unsrt}
\bibliography{ms.bbl}

\end{document}


\pagenumbering{arabic}
\date{}
\maketitle

\textbf{Supplementary Note 1 - Thermal boundary conductance measurements}

\begin{figure}[H]
\begin{center}
\includegraphics[width=\columnwidth]{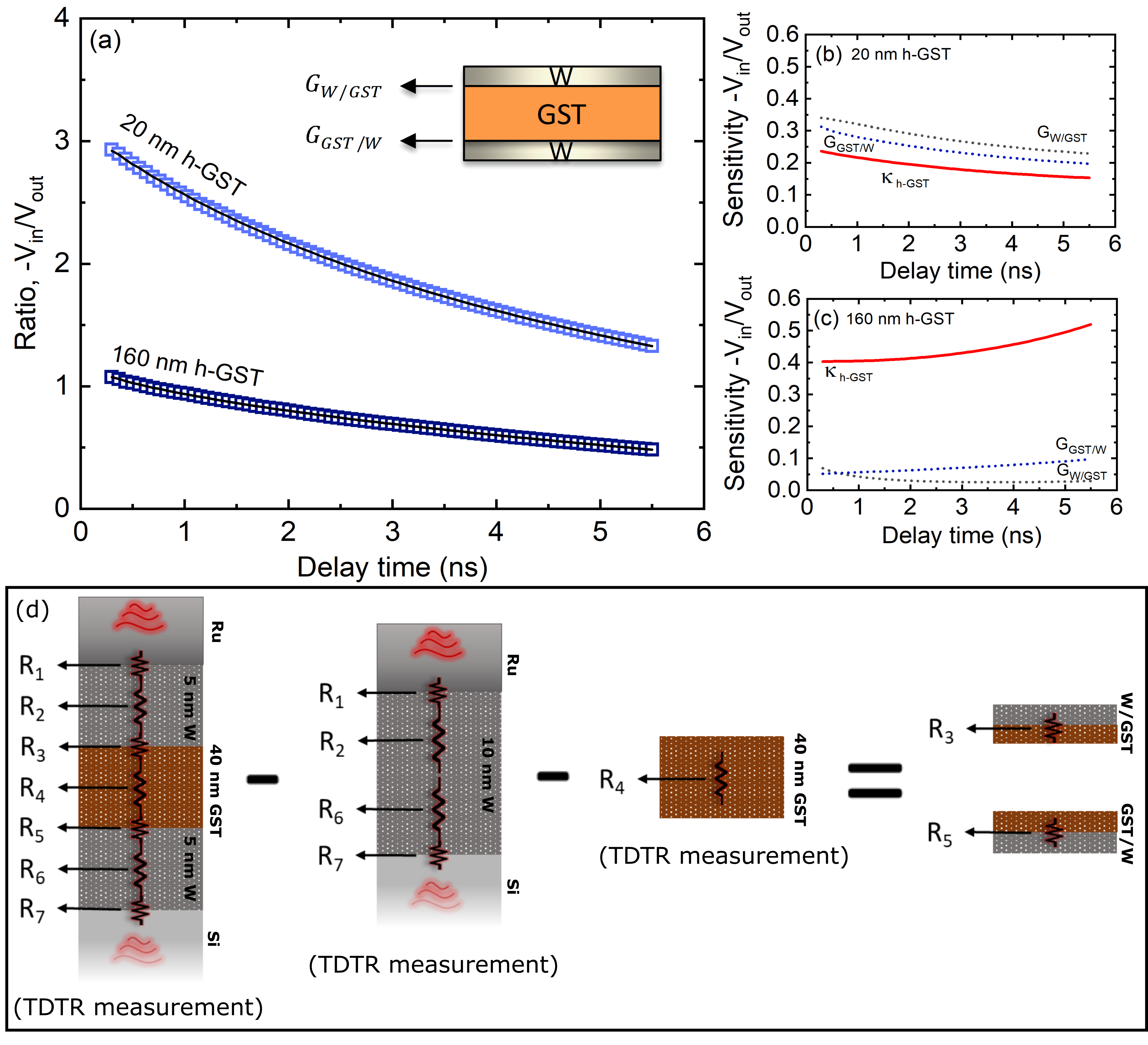}
\caption{(a) Theoretical fit for 20 and 160 nm GST thickness, (b,c) Sensitivity to thermal conductivity and TBC on either side of GST layer for a 20 and 160 nm GST (d) Schematic representing the approach used in this study to find the thermal boundary conductance between GST and W.
\label{fig:Fig_S1}}
\end{center}
\end{figure}

The thermal properties of Ge$_2$Sb$_2$Te$_4$ (GST) are measured via Time-domain Thermoreflectance (TDTR). The experimental procedure and the data analysis for this technique are extensively discussed elsewhere \cite{cahill2004analysis, schmidt2008pulse, hopkins2010criteria}. Figure \ref{fig:Fig_S1}(a) shows the experimental data with its corresponding theoretical fit for 20 and 160 nm h-GST. To find the GST thermal conductivity and the thermal boundary conductance (TBC) between GST and W (G$_{GST/W}$), we perform measurements on various thicknesses of GST. This is due to the fact that the sensitivity of our measurements to thermal conductivity and TBC varies with respect to thickness and, therefore, by measuring thermal conductance across various thicknesses we can distinguish the thermal conductivity from that of TBC. According to the sensitivity analysis in Fig. \ref{fig:Fig_S1} (b,c), for 20 nm GST, the sensitivity of our measurements to TBC is highest whereas in the 160 nm thick GST, the sensitivity to TBC is negligible. Therefore, we obtain the intrinsic thermal conductivity from 160 nm thick GST where any effect from TBC is minimum. 

Due to a lack of sufficient experimental sensitivity to TBC for amorphous and cubic phases, we are unable to directly measure TBC between GST and W. However, by performing a series of measurements targeting samples with different sensitivities to the parameters of interest, namely TBC and thermal conductivity, we can extract a reasonable approximation for TBC across different phases. For this, we measure the thermal conductance for a 20 and 40 nm sample, which includes the contribution of all resistances, i.e., layers and their corresponding interfaces. In the calculations of TBC, although it makes more sense to use the thinnest GST films like 5 and 10 nm, we refrained from taking them into account. The reason is because the thermal transport for these ultra-thin thicknesses are not fully within a diffusive regime. For diffusive thermal transport, the mean free paths of heat carriers must be shorter than the thickness of the film \cite{wilson2014anisotropic}. 

In 20 and 40 nm thick GST, the existence of a thin GST layer increases the sensitivity of our measurements to G$_{GST/W}$ (Fig. \ref{fig:Fig_S1}(b)). To find G$_{GST/W}$, as depicted in Fig. \ref{fig:Fig_S1}(d), we need to subtract the effect of all other resistors in the series from the total resistance. For this, since the 20 and 40 nm GST yield a relatively small resistance between the Ru and Si, we can treat the entire stack (W/GST/W) as an interface and using a two layer model, measure the thermal conductance (within the penetration depth) across the Ru/Si interface. The penetration depth in our measurement is on the order of $\sim$100 nm. The substrate is silicon which acts as a heat sink and therefore the resistance due to this layer is negligible. This leaves us with seven resistors between Ru and Si as depicted in Fig. \ref{fig:Fig_S1}(d).

The measured thermal conductance for 20 and 40 nm samples include the effect of interfaces and the intrinsic thermal conductivities of all the layers between Ru and Si. Now, in order to deconvolve the thermal conductivity from that of the TBC, we need to know the intrinsic thermal conductivity of each layer as well as their corresponding TBCs. For this, using a different set of samples, we measure the thermal conductance across Ru/10 nm W/Si to account for the intrinsic thermal conductivity of W, Ru/W, and W/Si interfaces. Next, assuming the intrinsic thermal conductivity of 20 and 40 nm thick GST film is similar to that of the 160 nm, we subtract the resistance due to the intrinsic thermal conductivity of the 20 and 40 nm GST layer from the total resistance. In order to mathematically derive an equation for estimating the TBC between GST and W, we assume each layer and interface introduces a resistance to the thermal transport from the transducer to the substrate similar to the schematic in Fig. \ref{fig:Fig_S1} (d). We can obtain the overall resistance of the stack between Ru and Si as follows:

\begin{equation}\label{eq1}
\begin{split}
R_{total} & = R_{W} + R_{GST} + R_{W} \\
& + R_{Ru/W} + R_{W/GST} + R_{GST/W} + R_{W/Si},
\end{split}
\end{equation}

Where R represents the thermal resistance and is defined as the inverse of conductance, R$_{total}$ = 1/G$_{total}$. Due to high thermal conductivity of W, the thermal resistance of the W layer compared to that of GST is essentially negligible and can be dropped from Eq \ref{eq1}. Additionally, due to electronic effect, the interfacial thermal resistance between metal-metal interface such as Ru/W is negligible. Furthermore, we assume the boundary conductance at the front and rear sides of the GST that are in contact with tungsten are identical ( $R_{W/GST} = R_{GST/W}$). As a result, Eq. \ref{eq1} can be simplified as follows:

\begin{equation} \label{eq2}
R_{\textrm{total}} = R_{GST} + 2R_{W/GST} + R_{W/Si},
\end{equation}

In the above equation, except for the R$_{W/GST}$, other parameters can be measured from TDTR and TEM. Considering that the thermal resistance is the inverse of the thermal conductance, by rearranging the terms in Eq. \ref{eq2} we can obtain an equation for GST/W TBC as follows:

\begin{equation} \label{eq4}
G_{\textrm{GST/W}} = \frac{ 2 } {(\frac{1}{G_{\textrm{total}}} - \frac{d_{film}}{k_{160 nm GST}} - \frac{1}{G_{\textrm{Ru/10 nm W/Si}}} ).} 
\end{equation}

\begin{figure}
\begin{center}
\includegraphics[width=\columnwidth]{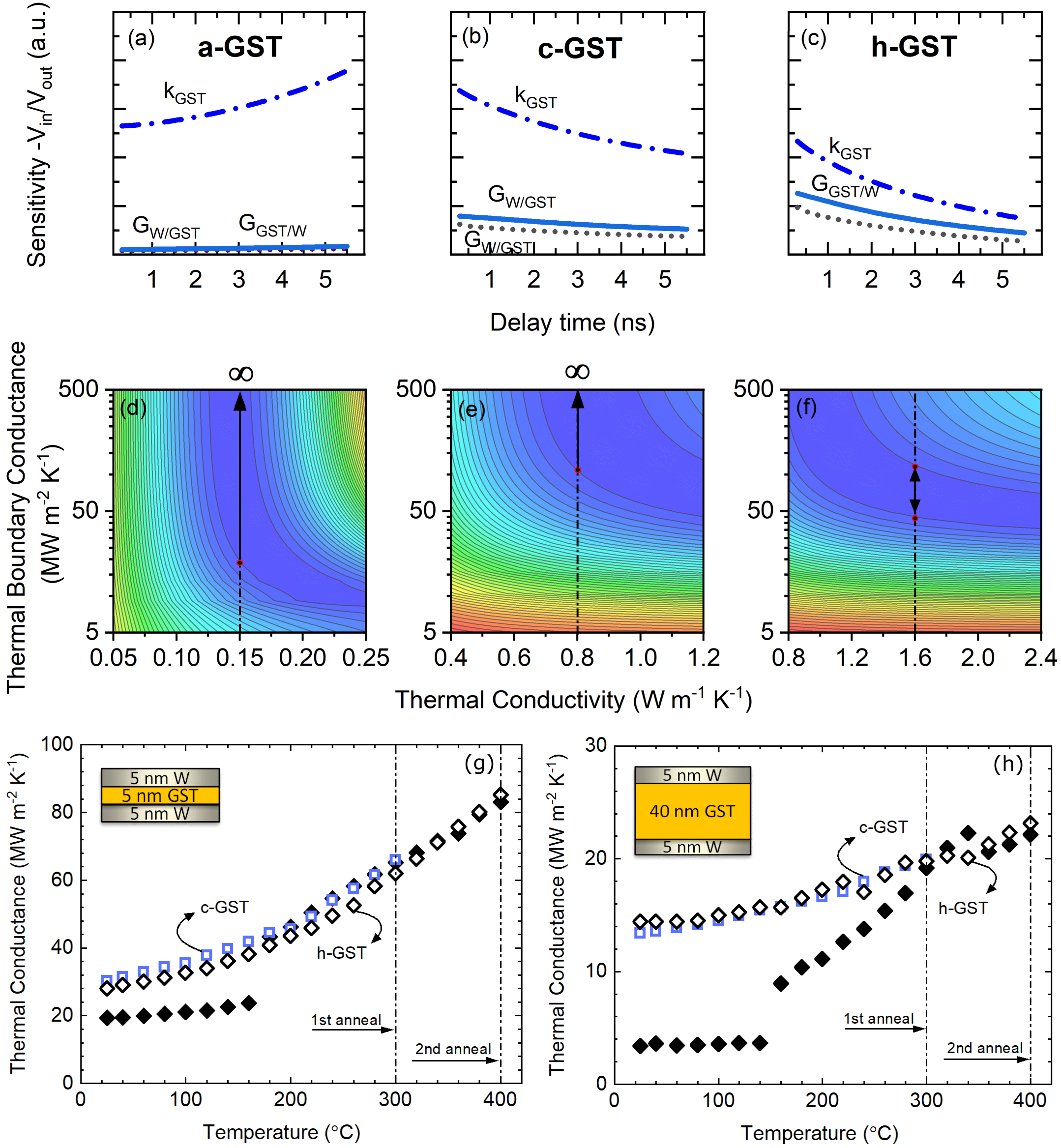}
\caption{ (a-c) Sensitivity of thermal conductivity and top/bottom interfaces in a 40 nm thick GST as a function of delay time for across phases. (d-f) Residual contour of our model's fit relative to the best fit for G$_{GST/W}$ as a function of thermal conductivity in different phases of GST. The solid line arrows indicate the range where for any G$_{GST/W}$ input, the model produces a good fit with correct thermal conductivity. (g,h) Thermal conductance for different thicknesses of GST, 5-10-20-40 nm, sandwiched between 2 and 5 nm W spacers, respectively.
\label{fig:Fig_S2}}
\end{center}
\end{figure}

In this analysis, G$_{\textrm{total}}$ is the total thermal conductance across the composite Ru/W/GST/W/Si stack and d$_{film}$ is the thickness of the GST film. Assuming a similar top and bottom interface between GST and W, we multiply the obtained TBC by two (in the numerator) to estimate an individual GST/W interface. Using Eq. 3, we provide an estimate for the G$_{W/GST}$ across different temperatures. However, due to low thermal conductivity of the GST in amorphous and cubic phases, the sensitivity to the G$_{W/GST}$ is not sufficient to enable us to directly fit for this parameter. Figure \ref{fig:Fig_S2} (a-c) shows the sensitivity of our measurements to parameters like thermal conductivity and TBC in each phases. There is negligible sensitivity to TBC in the amorphous phase. Upon transformation of a-GST to c-GST, the sensitivity to TBC increases but is still significantly lower than the measurement sensitivity to the thermal conductivity and therefore, would not affect the thermal conductivity measurements. The existence of low sensitivity to TBC in a-GST and c-GST, explains the similar trend observed for both 40 and 160 nm thick GST. However, as the c-GST transitions to h-GST, the sensitivity to the thermal conductivity decreases relative to those of the TBCs across the interfaces, and therefore, above 320 \degree C, the effective thermal conductivity is suppressed by the TBC.

Figure \ref{fig:Fig_S2} (a-c) explains our motivations behind using a lower bound in the main manuscript for G$_{W/GST}$ across different phases. This figure plots the residual value for our model's fit relative to the best-fit with respect to the input thermal conductivity and G$_{W/GST}$. For this, a range of different values for the thermal conductivity and G$_{W/GST}$ are used to fit our model to the experimental data, and based on the amount of deviation from the best-fit value these plots are generated for different phases of GST. The blue region in these plots corresponds to the minimum deviation from the best-fit value. In other words, for any thermal conductivity and G$_{W/GST}$ that are taken from the blue region, the model generates the same quality of fit to the empirical data. This being said, since we can find the thermal conductivity of GST using the thick sample (160 nm), technically, we should be able to directly fit for the G$_{W/GST}$ in our thermal model using the 20 or 40 nm GST measurements. However, as demonstrated in Fig. \ref{fig:Fig_S2} (d and e), due to lack of sensitivity in a-GST and c-GST, for any G$_{W/GST}$ that has a value higher than the red circle mark (20 and 110 MW m$^{-2}$ K$^{-1}$ for amorphous and cubic, respectively), the model produces a good fit. For example, in the amorphous phase, if we only fit for the thermal conductivity and fix G$_{W/GST}$ to any value between 20 MW m$^{-2}$ K$^{-1}$ and infinity, the model still produces a best-fit to the experimental data. On the other hand, in h-GST (\ref{fig:Fig_S2} (f)), the G$_{W/GST}$ that are in the range of 45-115 MW m$^{-2}$ K$^{-1}$ can produce a good fit. This observation leads to prescribing a minimum limit to  G$_{W/GST}$ for amorphous and cubic phases. In order to calculate a lower bound for the G$_{W/GST}$ across different phases, we assume 10$\%$ uncertainty in the measurement of parameters that play a role in the thermal transport such as thermal conductivity of GST and its thickness. The following table indicates the sign for calculation of uncertainty in our analysis. The sign is selected to ensure the lowest resulting G$_{W/GST}$.

\begin{table*}[h]
  \centering
  \begin{tabular}{@{}ccccc@{}}
    \toprule
    \toprule
    -  & G$_{tot}$  &  G$_{Ru/10 nm W/Si}$  & d$_{GST}$   & k$_{GST}$   \\
    \midrule
    $\ch{Uncertainty}$ & -10$\%$ & +10$\%$ & -10$\%$ & +10$\%$ \\
    \bottomrule
    \bottomrule
  \end{tabular}
\end{table*}

Similar to Fig. 3 (b) in the manuscript, in order to observe the thermal conductance of the stack when GST is in cubic vs.$\sim$ hexagonal phase, we perform a similar measurement for 5 and 40 nm thick films. For this, we initially measure the thermal conductance for an as-deposited GST across different temperatures to determine the thermal conductance as the phase transition occurs. Then, we take another as-deposited sample, anneal it to 300 \degree C, then cool the sample down to room temperature, and measure its room temperature thermal conductance while cooling. A similar sample, was annealed to 400 \degree C and the thermal conductance was measured as the sample cooled down. In this way, we obtained the thermal conductance across different temperatures for c-GST and h-GST. Figure \ref{fig:Fig_S2} (h,g) shows the results of these measurements. Annealing the 5 and 40 nm thick GST from 300 to 400 \degree C does not change the thermal conductance behaviour. This however, is surprising due to the fact that the thermal conductivity of GST for the annealed cases is 0.75 and 1.4 W m$^{-1}$ K$^{-1}$, respectively. This provides additional evidence for the reduction of TBC in the hexagonal phase as compared to cubic phase. We surmise that the reason we do not observe a pronounced difference between the c-GST and h-GST similar to Fig. 3(b) is because, in the case of 5 nm, the ballistic transport of carriers prevents us from observing the changes in the interfacial characteristics, and for the case of 40 nm, we are losing sensitivity to the TBC due to increased resistance of the GST film itself.

\begin{figure}
\begin{center}
\includegraphics[width=\columnwidth]{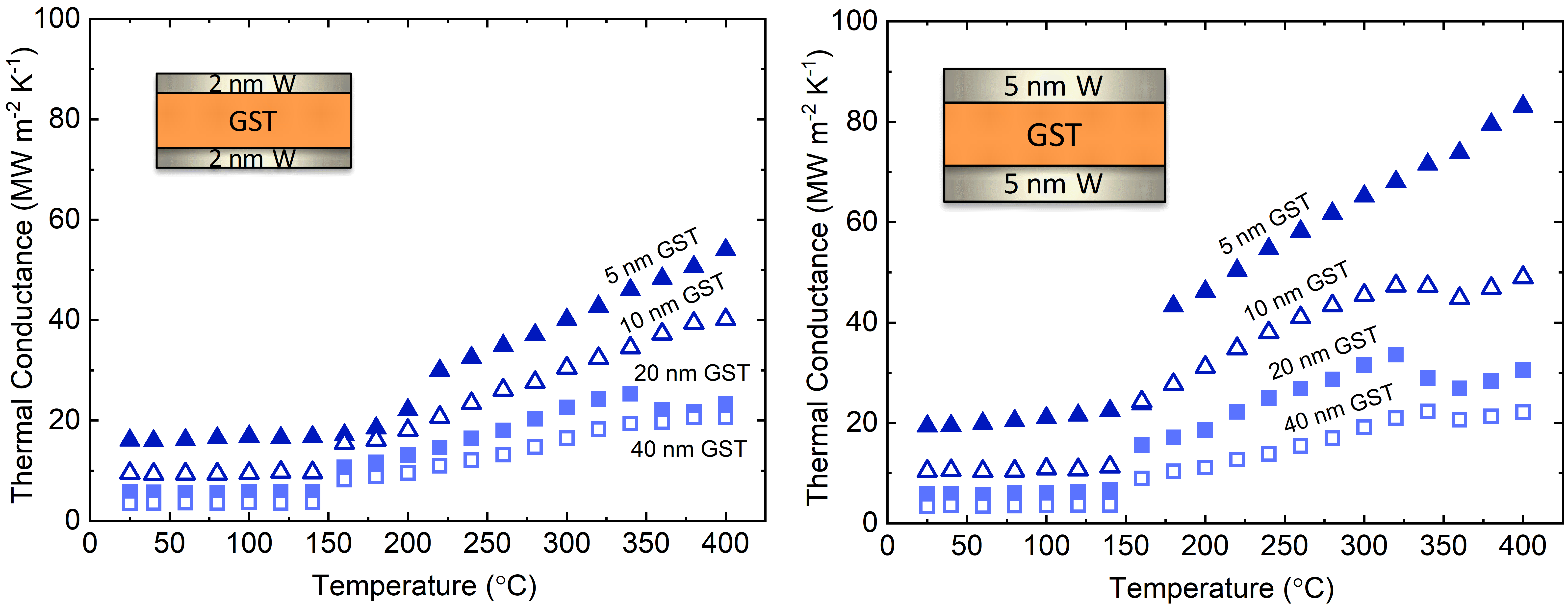}
\caption{Thermal conductance across Ru/W/GST/W/Si for different thicknesses of GST with (a) 5 nm and (b) 2 nm W spacers.
\label{fig:Fig_S3}}
\end{center}
\end{figure}

Figure \ref{fig:Fig_S3} (a,b) indicates the thermal conductance from room temperature to 400 \degree C for different thicknesses of GST sandwiched between 2 and 5 nm of W spacers, respectively. The thermal conductance is lower when the W spacer is 2 nm. This is also observed in the room temperature measurements as discussed in the main manuscript. Here, we observe a significant difference especially above the crystalline transition temperature ($\geq$ 150 \degree C). Above this temperature, the thermal conductance for 5 nm GST in both W thicknesses linearly increases with temperature. We attribute this to ballistic transport of phonon and electron across GST layer, as discussed in the main manuscript. It is worthwhile to mention that the effect of TBC between GST and W above 340 \degree C is most noticeable for GST thickness of 20 nm. This is because for thicknesses thinner than 20 nm, the thermal transport is not fully diffusive. Whereas, for thicknesses larger than 20 nm, the GST layer resistance dominates the thermal transport. These plots clearly indicate that utilizing the 2 nm W is a better choice for electrode as it results in a reduced thermal transport. We quantify this by coupling the experimental data with finite element simulations and find that merely reducing the W thickness can potentially lower the I$_{reset}$ by 5$\%$.

\begin{figure}
\begin{center}
\includegraphics[width=\columnwidth]{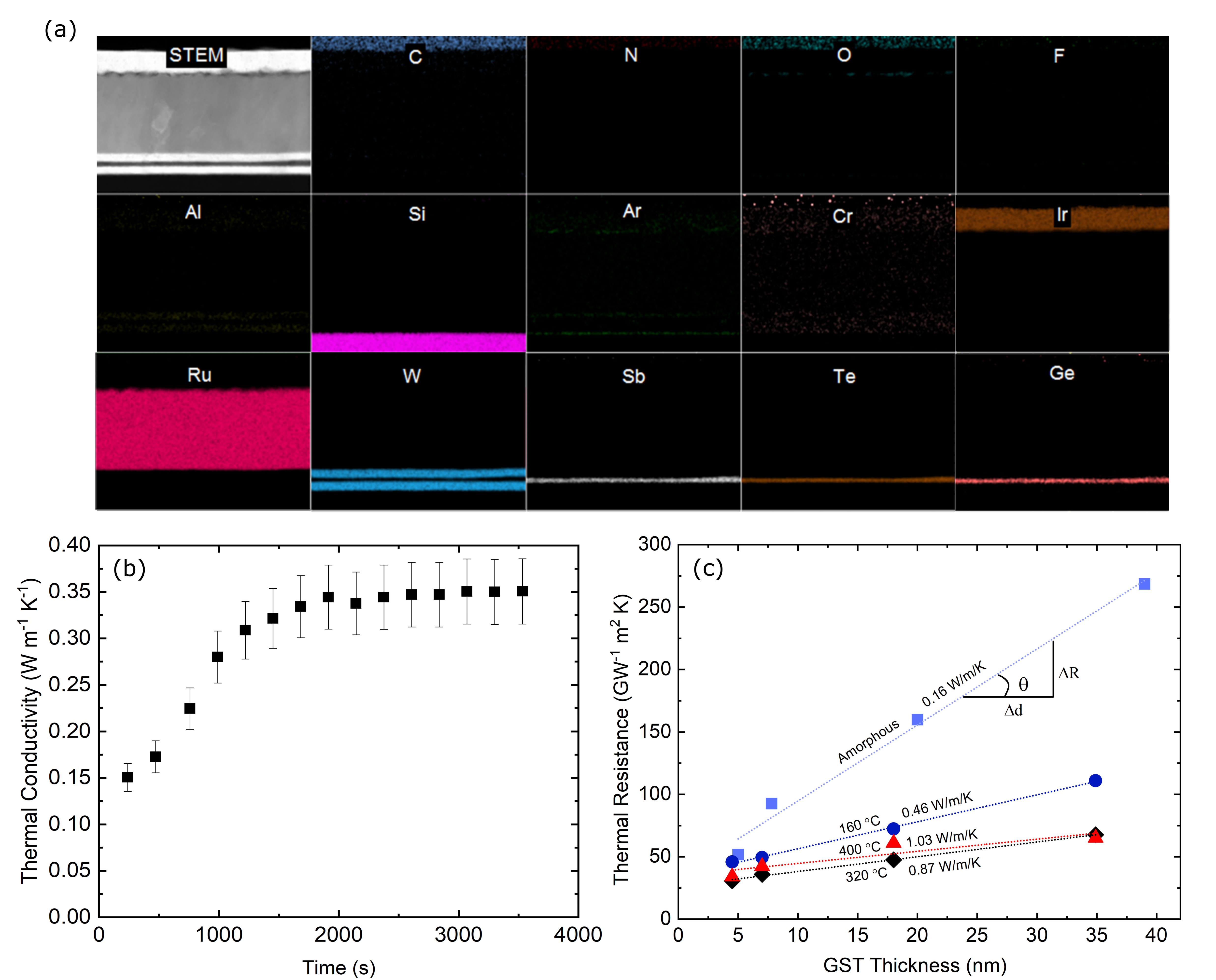}
\caption{(a) Elemental map for the constituents materials in the stack configuration studied (b) Thermal conductivity evolution as the GST transitions from amorphous to cubic crystalline at 140 \degree C as a function of time for a 40 nm thick GST film with W spacers. (c) Thermal conductivity of GST at different annealed temperature. Note, the thermal conductivity of 400 \degree C annealed case is measured lower than reported in the manuscript due to the effect of reduced thermal boundary conductance.
\label{fig:Fig_S5}}
\end{center}
\end{figure}

Figure \ref{fig:Fig_S5} (a) depicts the elemental mapping results for the GST layer determined from STEM, indicating the homogeneous distribution of Ge, Sb, and Te in a 5 nm GST layer. Figure \ref{fig:Fig_S5} (b) shows the TDTR measurement of the thermal conductivity as a function of time near the transition temperature (140 \degree C) for 40 nm GST. Prior to taking the measurements, the sample was heated up to 120 \degree C and after waiting long enough for the sample to equilibrate, the temperature was raised to 140 \degree C at a rate of 50 K/min. As can be seen, it takes nearly 2000 s for the GST layer to transition from amorphous to crystalline structure. Figure \ref{fig:Fig_S5} (c) indicates the thermal conductivity of thin GST films that have been annealed in a furnace for more than two hours at different temperatures. In our thin films, by applying a linear fit to the thermal resistance data as a function of the layer thickness, the thermal conductivity can be found. This method, however, only applies to GST for amorphous and cubic phase where the sensitivity to TBC is negligible, and we do not expect size effects in the thermal conductivity due to the relatively small mean free paths in the GST in these phases. As can be seen, the thermal conductivity increases as the annealing temperature increases and the values match the those obtained from the 160 nm measurements. On the other hand, in the hexagonal phase, since the reduction in TBC influences the measurements, the obtained resistance for the sample that is annealed to 400 \degree C is close or even higher than the 320 \degree C sample. Considering that h-GST has a factor of two higher thermal conductivity than c-GST, the measurement of higher thermal resistance for 400 \degree C annealed samples than 320 \degree C is unexpected. After performing \textit{in situ} TEM and confirming that the GST film has not been damaged due to heating, we attributed this to the change in the thermal boundary conductance of GST with W layer.

\newpage
\textbf{Supplementary Note 2 - Electron vs. phonon contribution in thermal conductivity} 

\begin{table}
  \caption{Electrical resistivity for GST measured by different groups \cite{kato2005electronic, nirschl2007write,lee2013phonon, siegrist2011disorder, lyeo2006thermal,bragaglia2016metal} and the corresponding thermal conductivity calculated from WF.}
  \label{tbl:elect_table}
  \includegraphics[width=\linewidth]{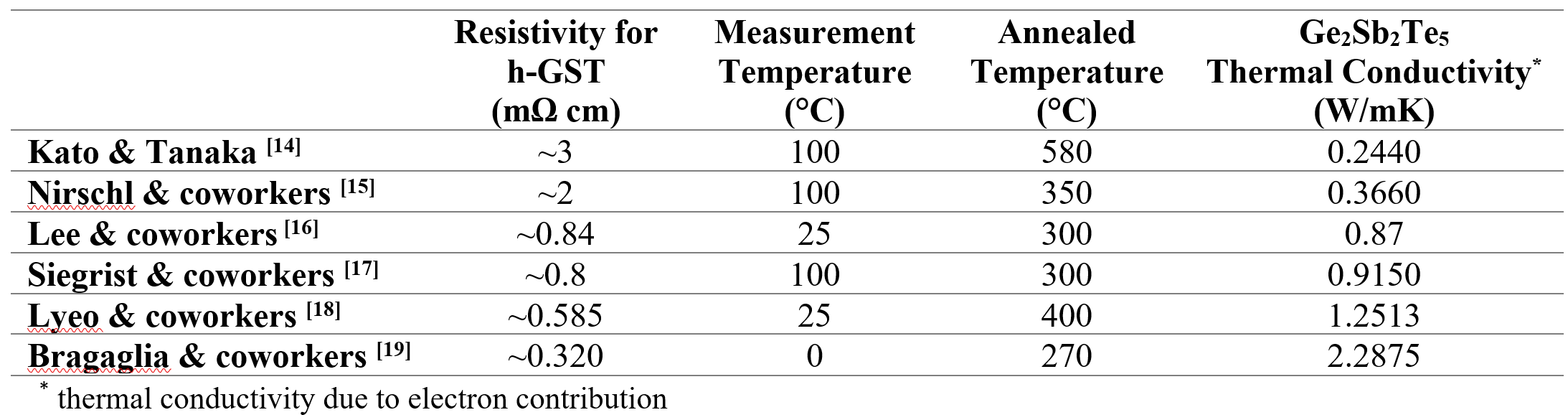}
\end{table}

An important factor in the thermal transport mechanism in GST is the contribution from the electrons vs. phonons in the total thermal conductivity. Application of the Wiedmann-Franz (WF) law is a common approach that makes use of electrical resistivity for estimating the electronic contribution in thermal conductivity.

\begin{equation}
   k = k_p + k_e
\end{equation}

\begin{equation}
   k_e=  LT/\rho
\end{equation}

\noindent
where k$_p$  and k$_e$ are thermal conductivities due to phonon and electron contribution, respectively, L is the Lorenz number, often assumed as the low temperature value of 2.44 $\times$ 10$^{-8}$ W $\Omega$ K$^{-2}$, T is temperature, and $\rho$ is the electrical resistivity. For example, Lyeo and coworkers \cite{lyeo2006thermal} reported negligible electronic contribution in \textit{a}-GST and c-GST, while 70\% contribution in h-GST based on electrical resistivity measurements. However, a survey of the data available in literature, as given in Table \ref{tbl:elect_table} for the electrical resistivity of the h-GST reveals a significant variations among reported values for the electrical resistivity of h-GST, ranging by as much as an order of magnitude. This difference among the electrical resistivities in different studies could be partially due to the different deposition process, composition variation, annealing time, or different measurement techniques. For example, Bragaglia \textit{et al.} \cite{bragaglia2016metal} reported that the resistivity of the h-GST largely depends on the degree of order in vacancy layers. They showed that for single crystalline h-GST, where the vacancy layers are highly ordered, the electrical resistivity could be substantially lower than reported values.

According to these studies, in h-GST, depending on the degree of disorder the thermal conductivity can largely vary. This is consistent with the observation of a disorder-induced metal-insulator transition in h-GST \cite{siegrist2011disorder}. However, as the system transitions towards more order, as well as increased electron thermal conductivity the lattice thermal conductivity is expected to increase. First principle calculations demonstrate that the lattice thermal conductivity of bulk h-GST can vary in the range of 0.87-1.67 W m$^{-1}$ K$^{-1}$ depending on the crystal orientation \cite{mukhopadhyay2016optic}. Similarly, using first principle calculations, Campi et al. \cite{campi2017first} showed that by adding various scattering terms (Sb/Ge sublattice disorder and vacancies), the lattice thermal conductivity of bulk h-GST can be adjusted to reduce from ideal value of $\sim$1.6 W m$^{-1}$ K$^{-1}$ to experimentally reported value of $\sim$0.45 W m$^{-1}$ K$^{-1}$. Perhaps, a focused study on thermal conductivity of h-GST at the metal-insulator transition would address this question. However, such study is beyond the scope of current paper.

\newpage
\textbf{Supplementary Note 3 - Why does interfacial resistance between GST and W change as the GST transitions from cubic to hexagonal?}

As the GST undergoes the cubic to hexagonal phase transition, not only does the lattice structure change, but so do the electronic structure and the bonding. Obviously, the variation of several properties in GST makes it exceedingly difficult to pinpoint the exact reasons behind the observed reduction in TBC. Nonetheless, in order to provide more insight into the role of crystal structure and interfacial disorder on the observed transition in TBC, we conduct a series of molecular dynamics simulations of the TBC across cubic and hexagonal close packed interfaces of materials that have equivalent masses to W and GST using a 6-12 Lennard-Jones (LJ) potential. Lennard-Jones is a 2-body potential; therefore, the only free parameter is the distance between the atoms, and this enables us to create different lattice structures using the same potential. This therefore allows us to study the role of crystal structure and disorder on TBC without making any assumptions regarding changes in the bonding character from the cubic to hexagonal phases. To this extent, this highlights the advantages of conducting these molecular dynamics simulations using the LJ potential. Additionally, the simplicity of these potentials allows us to assess our hypotheses to general classes of materials, thus providing means to broadly study our posits of the origin of reduction in TBC across the crystalline phase transitions.

\begin{figure}[H]
\begin{center}
\includegraphics[width=\columnwidth]{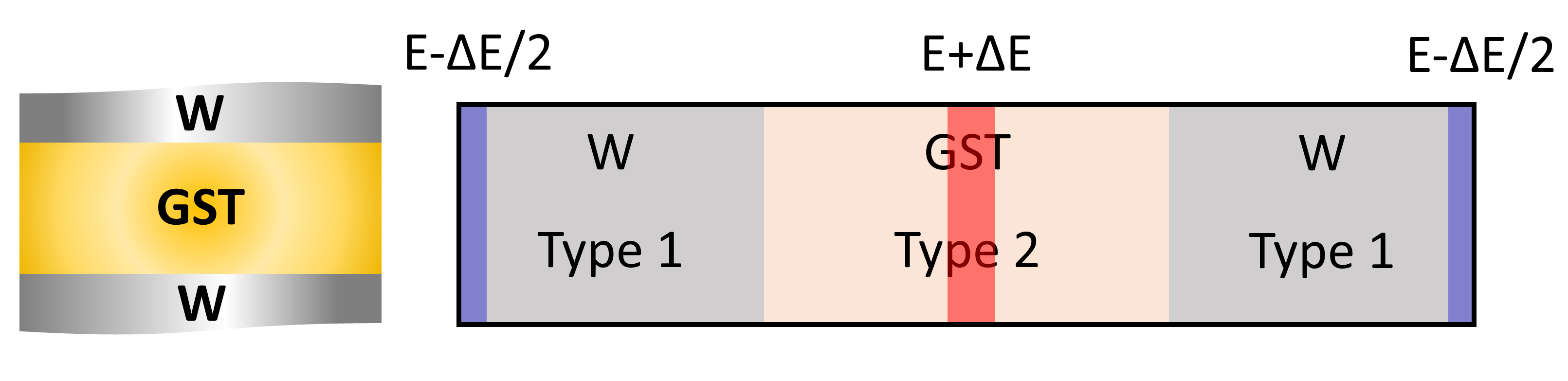}
\caption{Simulation set up and the location of heat baths for the molecular dynamic calculations.
\label{fig:Fig_SX1}}
\end{center}
\end{figure}

Since we are using LJ potentials to describe the crystalline W and GST films, in order to avoid confusion or misrepresentation, we call the section that represents W as \textit{type 1} and the section that represents GST as \textit{type 2}. With that in mind, we use parameters provided by Filippova et al. \cite{filippova2015calculation} for solid tungsten at room temperature ($\epsilon$= 1.451420 eV and $\sigma$ = 2.50374 nm). Although according to the paper, these parameters are supposed to result in a BCC lattice structure, we observe the lattice is unstable and tends to reorient to FCC structure. Nonetheless, we use this potential since our main purpose here is to investigate the effect of structural \textit{changes} on TBC. For the atoms in type 2 (GST), we could use a LJ potential with softer bonding energy compared to that of tungsten, yet, to keep the model as simple as possible, we use the same potential across all atom types in the GST. This allows us to only survey the effect of changes in the lattice structure. Thus, the only parameters that are different between type 1 (W) and type 2 (GST) are average atomic masses and number density. The atomic mass for type 1 is similar to that of W (183.84 u) and for type 2 is the arithmetic average of Ge$_2$Sb$_2$Te$_4$ (112.4 u). The number density for W and GST in our model is calculated to be $\sim$6.6 $\times$ 10$^{28}$ m$^{-3}$ and $\sim$ 2.7 $\times$ 10$^{28}$ m$^{-3}$ which stays relatively constant across all phases. For computational efficiency, a cutoff distance of 5.5 $\textrm{\AA}$ is used. For estimating the TBC at the interface between type 1 and type 2, we use a simulation box of 300 $\textrm{\AA}$ length with cross section area of 50$\times$50 $\textrm{\AA}^2$. In order to investigate the effect of disorder at the interface, we used a melt-quench technique to amorphize type 2 (GST) atoms. However, due to ordered interface of type 1, the amorphous structure nucleates near interface and turns into a thin FCC layer at the type-1/type-2 interface. We refer to this nucleated region as a disordered crystalline region which shows a higher TBC as compared to our ``ordered'' crystalline interfaces. The summary of our calculated TBC between different lattice structure are presented in table \ref{tbl:TBC_table}:

\begin{table}[H]
  \caption{Thermal boundary conductance (TBC) and resistance (TBR) across the interface between different lattice structures.}
  \label{tbl:TBC_table}
  \includegraphics[width=\linewidth]{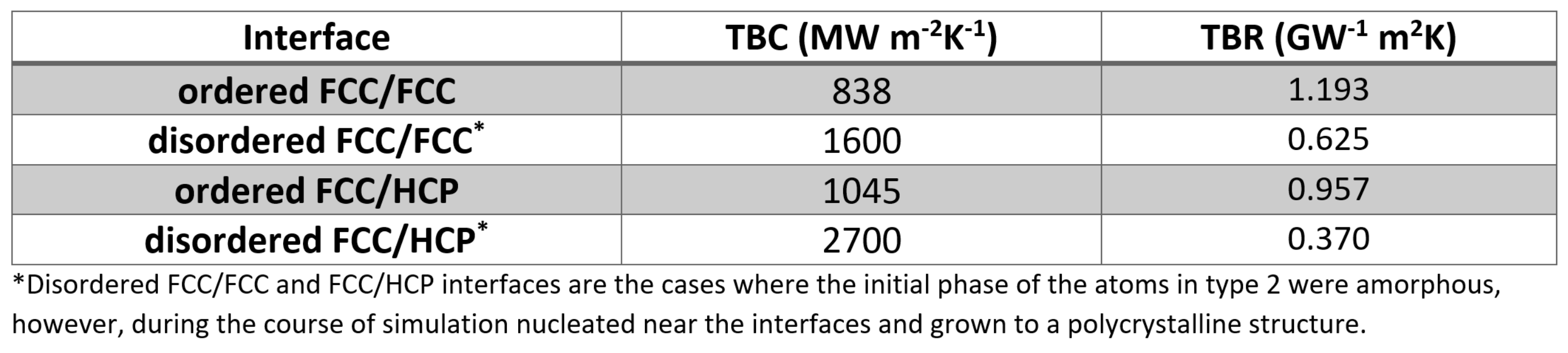}
\end{table}

Our results suggest that a change in phase from cubic to HCP does not significantly change the thermal boundary conductance. However, structural disorder at the interface could play an important role in the reduction of TBC from the cubic to HCP phase in our measured data across the W/GST/W interfaces. This is consistent with previous computational and experimental observations regarding the effect of disorder at the interface on the enhancement of TBC \cite{tian2012enhancing, english2012enhancing, gorham2014ion, giri2020review}. Tian et al. \cite{tian2012enhancing} used a theoretical approach - atomistic Green's function- and showed that the interface roughness in Si/Ge can increase phonon transmission compared to an ideal sharp interface. They concluded that this effect is even more pronounced if the acoustic mismatch between the materials at the interface is large, which is the case for GST and W. Several molecular dynamics simulations \cite{english2012enhancing, giri2015kapitza} have shown that compositionally disordered interfaces show higher TBCs than sharp interfaces. In addition, Gorham et al. \cite{gorham2014ion} experimentally showed that TBC can increase across ion irradiated interfaces of Al/native oxide/Si with sufficiently high ion dose due to compositional mixing and point defect formation. With respect to these previous works on the effect of disorder at the interface supported by our MD simulations, we hypothesize that one driving factor for the reduction in TBC from cubic to hexagonal phase could be due to the reduction of disorder rather than structural phase transition.

\begin{figure}
\begin{center}
\includegraphics[scale = 1.2]{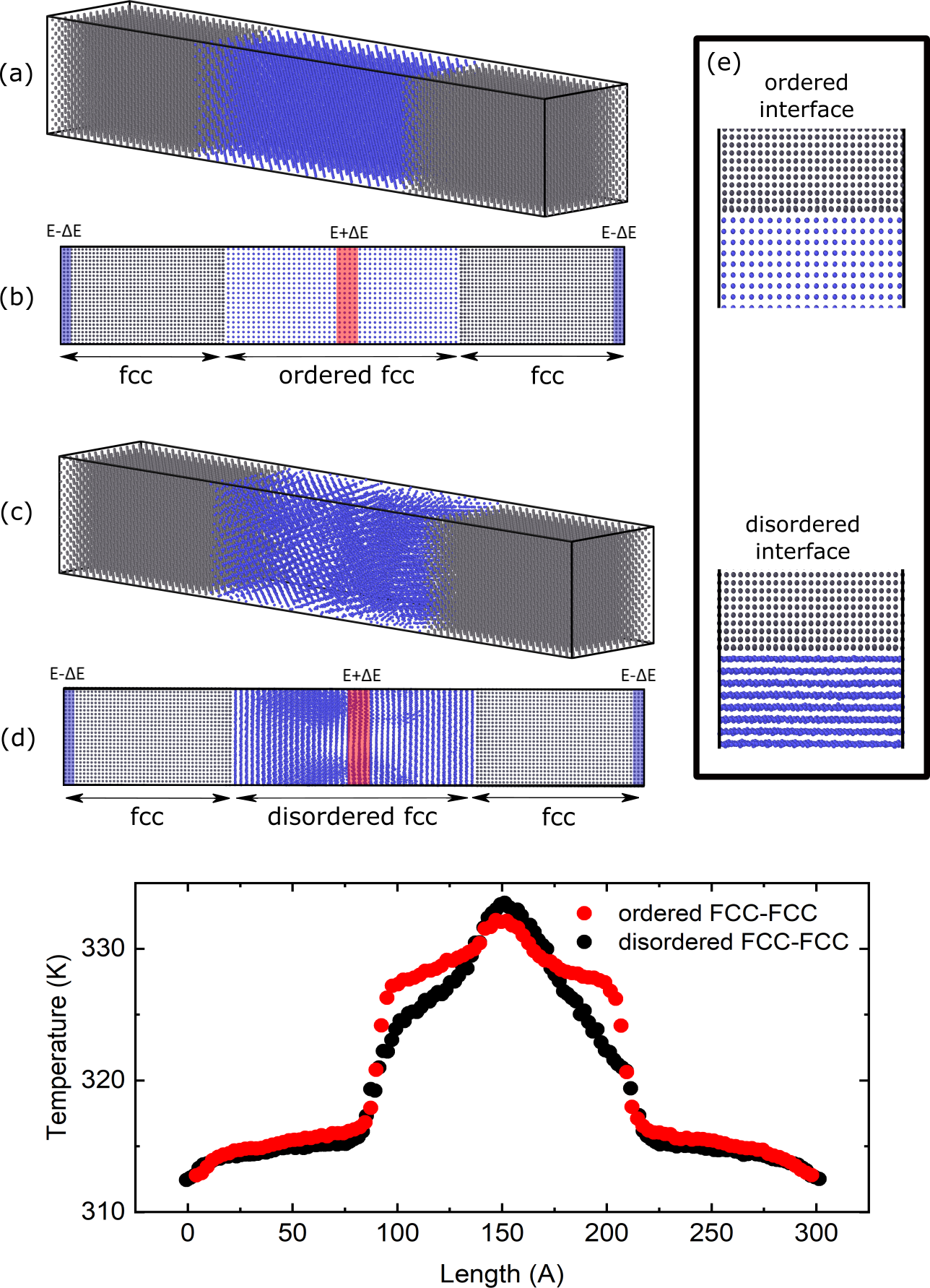}
\caption{Molecular dynamics simulation results for the system size of 50$\times$50$\times$300 $\textrm{\AA}^3$. (a,b) 3D and 2D visualization of the atomic arrangement in  the simulation after 6 ns for cubic/cubic/cubic structure. (c,d) 3D and 2D visualization of the atomic arrangement in  the simulation after 6 ns for cubic/disordered cubic/cubic structure. The disordered cubic phase is the result of nucleation from an amorphous phase. (e) The quality of interface after 6 million timesteps for interfaces with different quality. (f) Temperature profile along the simulation box when $\Delta$E = 1.5 eV/ps is added and subtracted from the hot and cold region depicted in red and blue (b,e). We calculate the TBC to be 838 MW m$^{-2}$ K$^{-1}$ and 1600 MW m$^{-2}$ K$^{-1}$ for ordered fcc/fcc and disordered fcc/fcc interfaces.
\label{fig:Fig_SX2}}
\end{center}
\end{figure}

\begin{figure}
\begin{center}
\includegraphics[scale = 1.2]{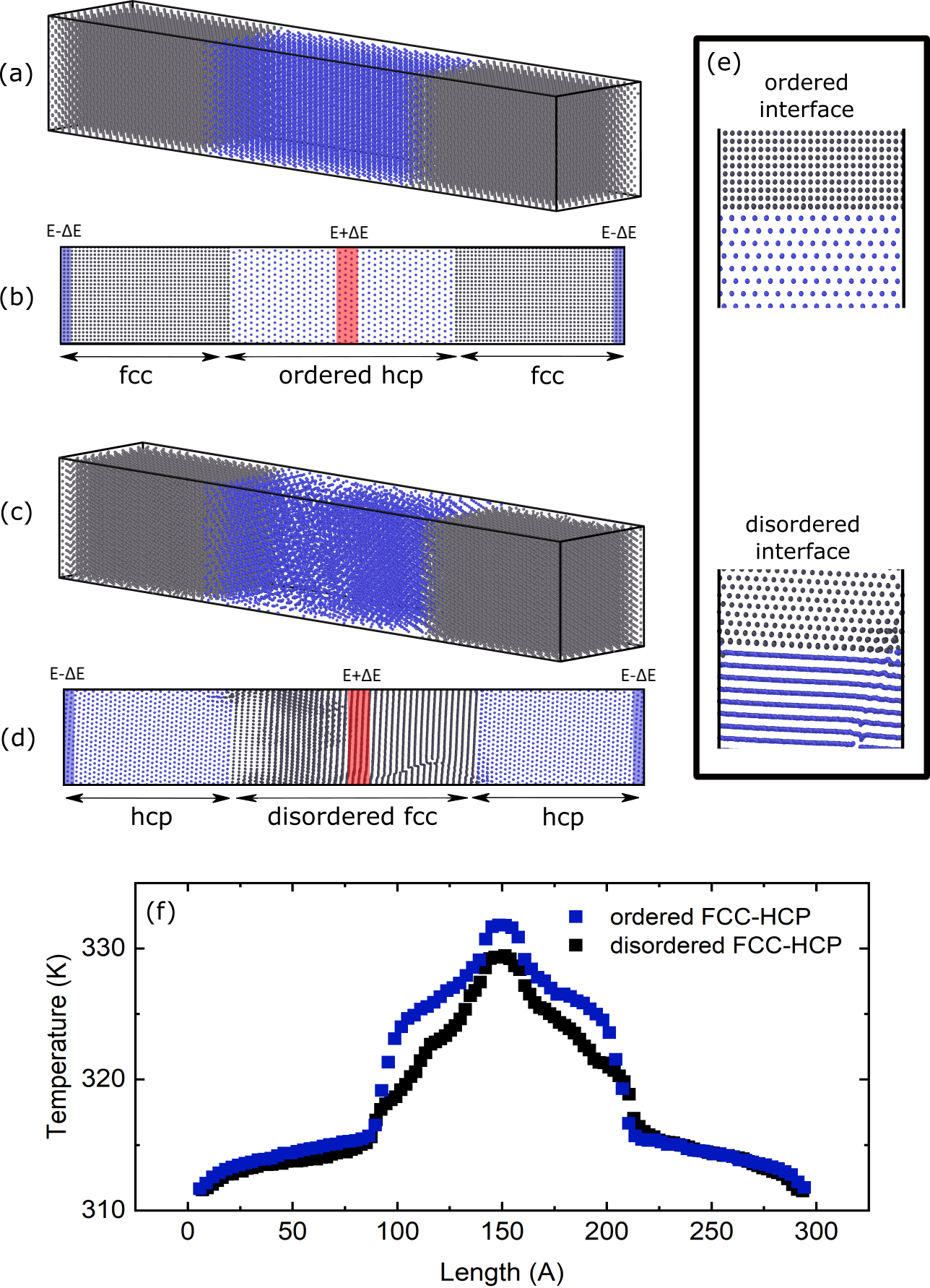}
\caption{Molecular dynamics simulation results for the system size of 50$\times$50$\times$300 $\textrm{\AA}^3$. (a,b) 3D and 2D visualization of the atomic arrangement in  the simulation after 6 ns for cubic/hexagonal/cubic structure. (c,d) 3D and 2D visualization of the atomic arrangement in  the simulation after 6 ns for hexagonal/disordered fcc/hexagonal structure. The disordered cubic phase is the result of nucleation from an amorphous phase. (e) The quality of interface after 6 million timesteps for interfaces with different quality. (f) Temperature profile along the simulation box when $\Delta$E = 1.5 eV/ps is added and subtracted from the hot and cold region depicted in red and blue (b,e). We calculate the TBC to be 1045 MW m$^{-2}$ K$^{-1}$ and 2700 MW m$^{-2}$ K$^{-1}$ for ordered fcc/hcp and disordered fcc/hcp interfaces.
\label{fig:Fig_SX3}}
\end{center}
\end{figure}

\begin{figure}
\begin{center}
\includegraphics[width=\columnwidth]{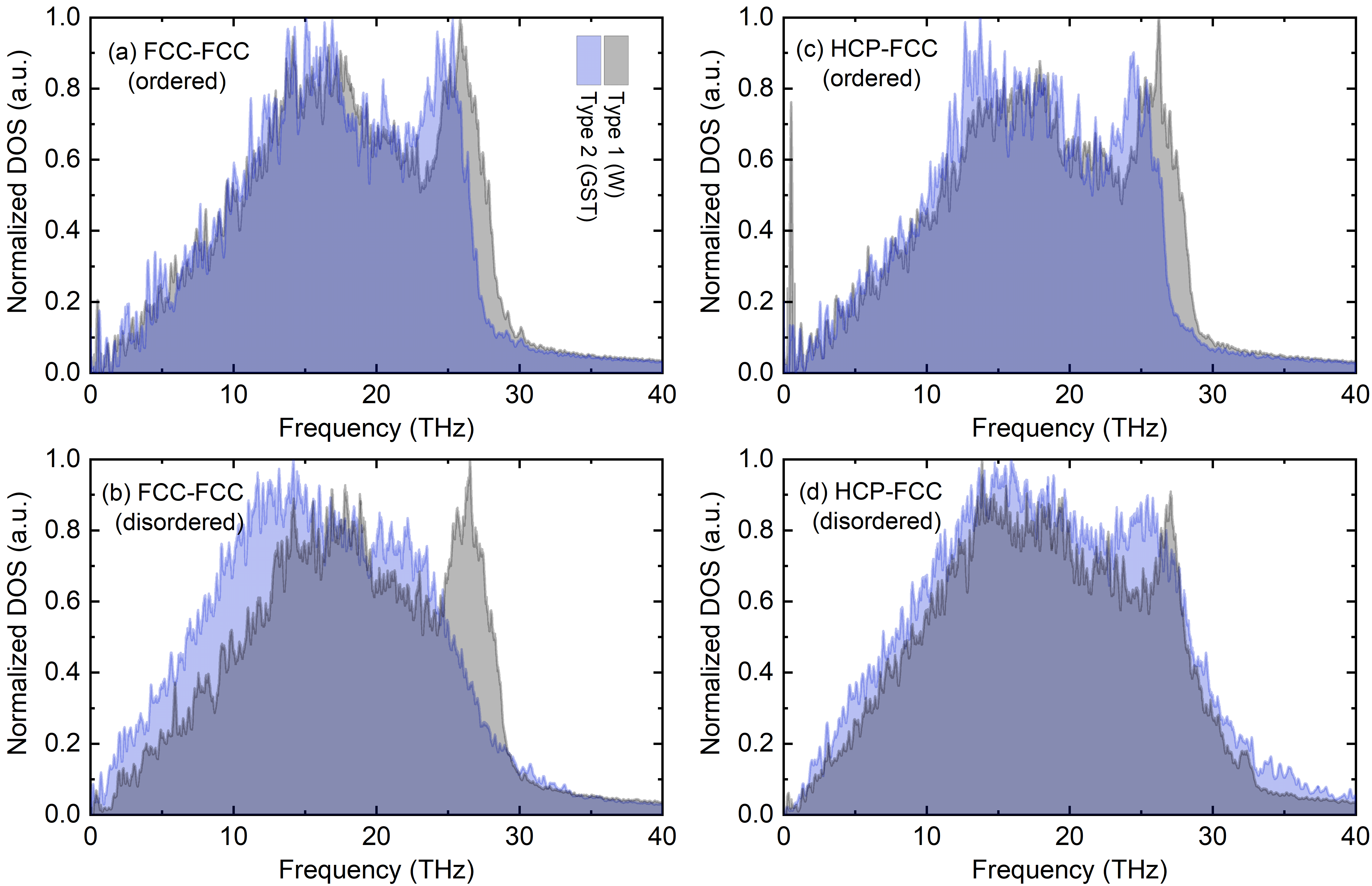}
\caption{Normalized density of states for type 1 and type 2 for different cases studied here.
\label{fig:Fig_SX4}}
\end{center}
\end{figure}

\newpage
\textbf{Supplementary Note 4 - Sound speed measurement in ultra-thin GST}

\begin{figure}[H]
\begin{center}
\includegraphics[width=\columnwidth]{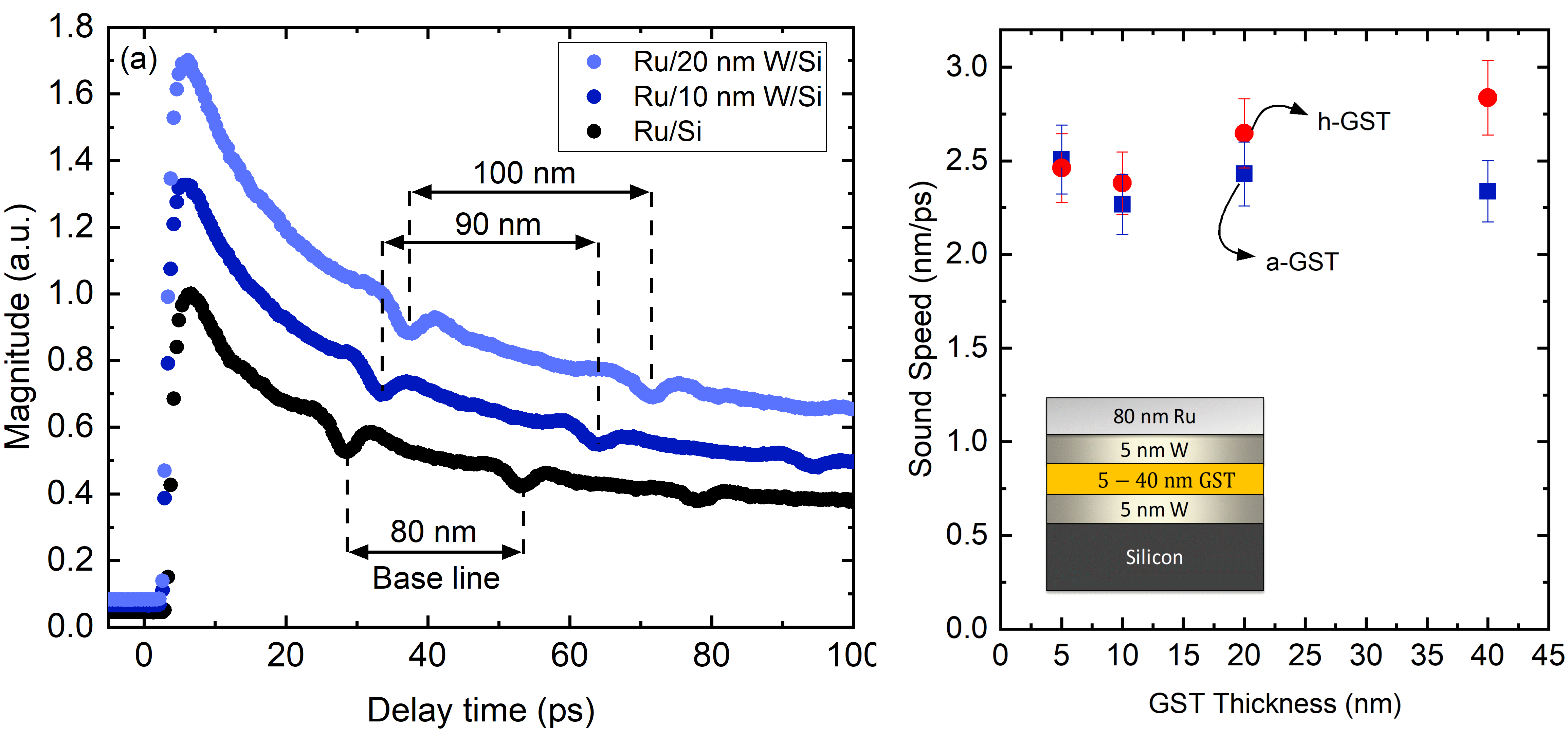}
\caption{Picosecond ultrasonic measurement in absence and presence of W layer.
\label{fig:Fig_S7}}
\end{center}
\end{figure}

The phonon mean free path in materials plays an important role in the analysis of thermal conductivity which can be estimated with the knowledge of the sound speed, specific heat, and thermal conductivity \cite{battaglia2010thermal}. Here, using the picosecond ultrasonic technique, we estimate the sound speed in GST for thicknesses less than 40 nm. In the configuration studied here, due to incorporation of multiple thin layers on top of each other (inset in Fig. \ref{fig:Fig_S7} (b)), the interpretation of picosecond ultrasonic data can be complicated by the existence of reflections off different interfaces. Therefore, in order to decipher the picosecond ultrasonic results accurately, we began our measurements with a simple substrate/transducer sample and gradually added more layers to the stack to deconvolve the effects of additional layers on the picosecond ultrasonic signals. Figure \ref{fig:Fig_S7} (a) shows the time evolution of strain waves travelling through ruthenium transducers with varying thickness of tungsten interlayers on silicon substrate via picosecond ultrasonic measurements. As can be clearly seen in the Fig. \ref{fig:Fig_S7} (a), every time a strain wave reflects off an interface and returns to the surface, a dip appears in the resultant thermoreflectivity decay curve which can be used to measure the exact thickness of each layer with knowledge of the sound speed in the material layers. The first decay curve corresponds to the base line where by measuring the time span between the echoes, assuming a sound speed of 6150 m/s for ruthenium, we can calculate the distance travelled for each wave which corresponds to the thickness of the layers. With the addition of 10 and 20 nm of tungsten between the ruthenium and silicon, one would expect to observe a secondary dip in the decay curve owing to the incorporation of an additional interface. However, due to the relatively low acoustic mismatch at the Ru/W interface, most of the wave passes through the Ru/W interface with no significant reflection. In practice, the addition of tungsten layer manifests itself as an increase in the distance between the echoes in the the thermal decay curve. In other words, the interface between Ru and W does not have a noticeable impact on the transmission of strain waves and they are only reflected off of the W/Si interface. In the light of this result, we conclude the Ru/W interface will not affect the picosecond acoustic signal which simplifies the interpretation of data when a GST layer is added to the system.

\begin{figure}[H]
\begin{center}
\includegraphics[scale = 0.5]{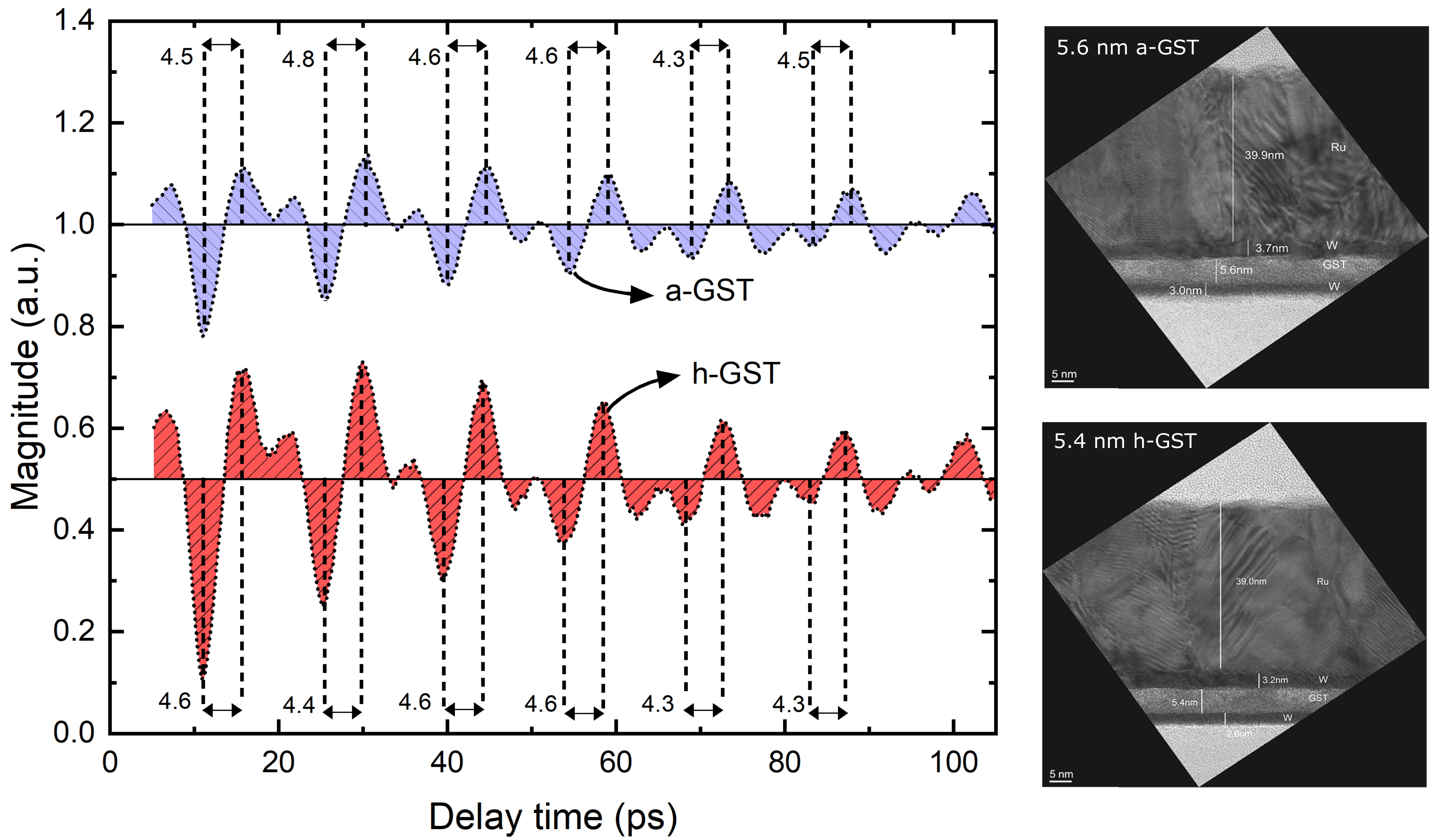}
\caption{Residual plot for picosecond ultrasonic measurement of a 5 nm thick GST in amorphous and hexagonal phase and the corresponding travelling time of strain waves in the GST layer.
\label{fig:Fig_S8}}
\end{center}
\end{figure}

\begin{figure}[H]
\begin{center}
\includegraphics[width=\columnwidth]{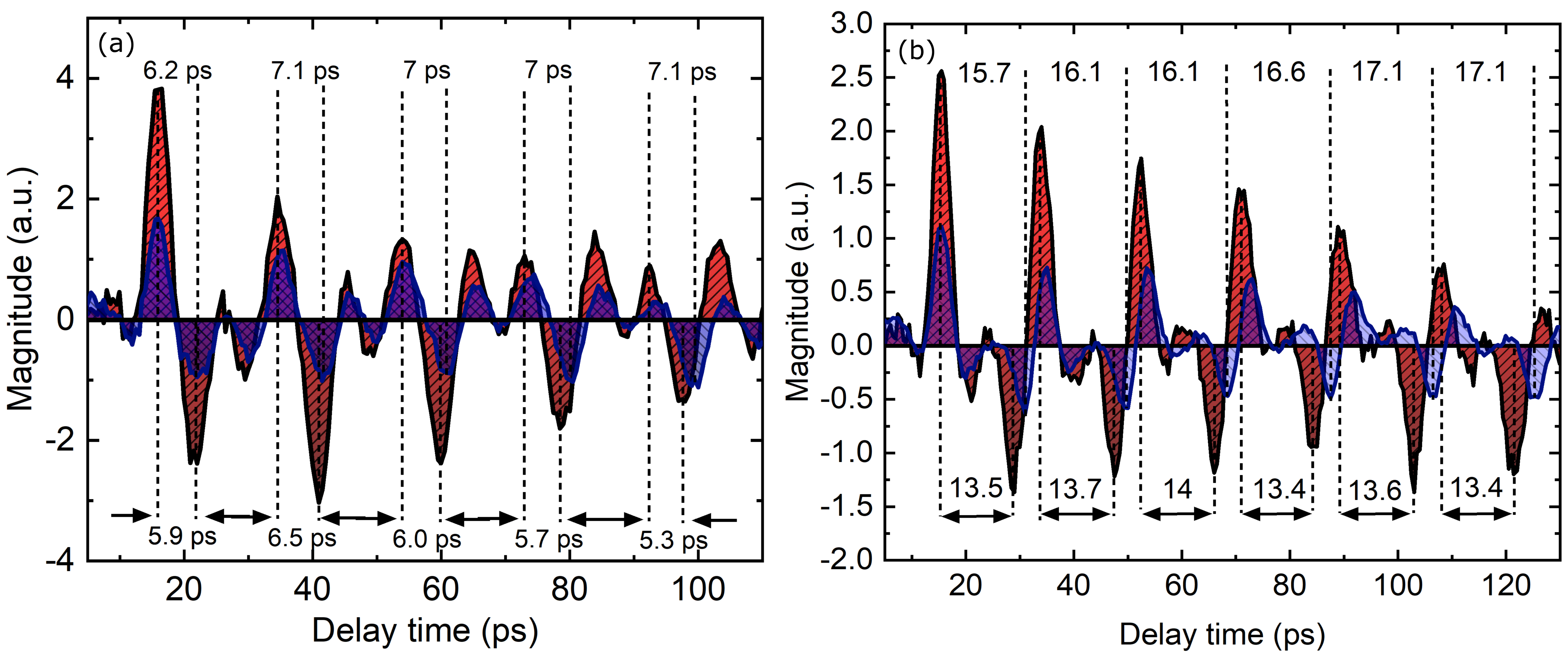}
\caption{Residual plot for picosecond ultrasonic measurement of (a) 10 nm and (b) 20 nm thick GST in amorphous and hexagonal phase and the corresponding travelling time of strain waves in the GST layer.
\label{fig:Fig_S9}}
\end{center}
\end{figure}

\newpage
\textbf{Supplementary Note 5 - Finite Element Simulation}
We model the propagation of strain waves using finite element (FE) simulations to ensure that our interpretation of the picosecond ultrasonic echoes are correct. In these simulations, the density, Poisson ratio, and the longitudinal sound speed are used as an input to determine the location of echoes in time. As such, the solid lines in Fig. 5 (a) in the main manuscript correspond to the simulation results and the dotted line corresponds to the picosecond residuals. As can be seen in Fig. 5 (a), the “humps” and “troughs” in the residual plots agree well with the simulations. The agreement for the location of the echoes between the simulations and the experiment confirms that they are not an artifact of measurement and are directly related to the reflection of the strain waves from the interfaces. The schematic in Fig. 5 (b) depicts the propagation of strain waves across different layers for the configuration studied here. As can be seen in Fig 5 (b) $i$, a strain wave is launched from the surface and travels across the Ru layer. Upon reaching the Ru/W interface, a lack of sufficient acoustic mismatch between Ru and W, allows the wave packet to completely pass through the interface without any interference (Fig 5 (b) $ii$). On the other hand, once the strain wave reaches the W/GST interface (Fig 5 (b) $iii$), as a result of large acoustic mismatch between W and GST, the wave is partially reflected and travels back to the surface and appears as upward “humps” in the residual plot. The other portion of the wave that passes the interface travels across the GST layer, and again, is partially reflected upon reaching the other GST/W interface where the consequence of this reflection appears as downward “troughs” in the residual plot. Using the time it takes for the strain waves to travel across the film, we can estimate the longitudinal sound speed.

\begin{figure}
\begin{center}
\includegraphics[width=\columnwidth]{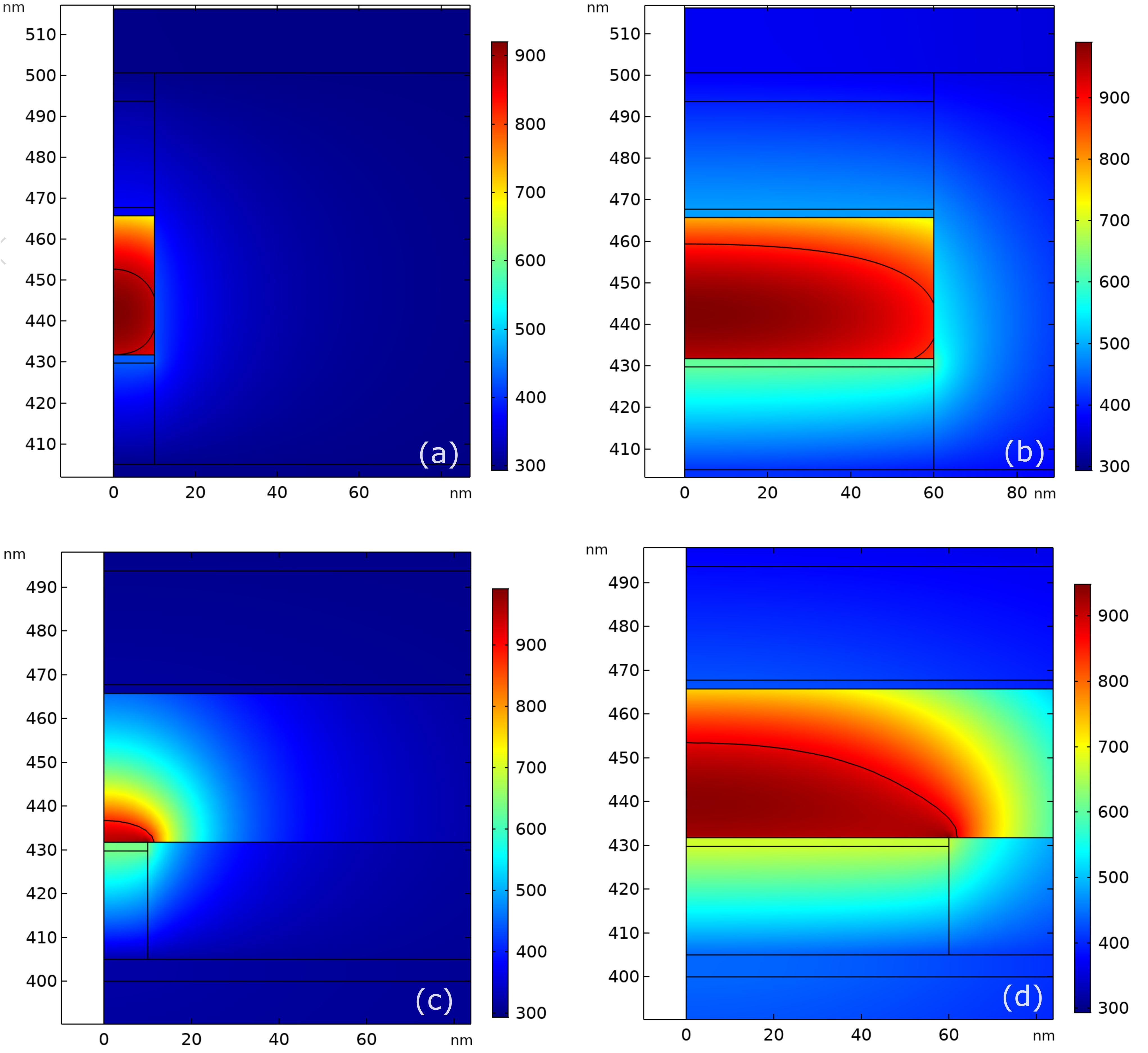}
\caption{ (a,b) Temperature profile for a 35 nm thick GST memory cell with a confined cell geometry with 20 and 120 nm lateral size. (c,d) Temperature profile for a 35 nm thick GST memory cell with a mushroom cell geometry with 20 and 120 nm lateral size.
\label{fig:Fig_S4}}
\end{center}
\end{figure}

Figures \ref{fig:Fig_S4} (a) and (b) depict the temperature profiles for a typical confined cell PCM device of 20 nm and 120 nm diameter, respectively, when the devices are subject to I$_{reset}$ and reach thermal equilibrium. The I$_{reset}$ is determined by ramping up the applied current until the 880 K isothermal lines (black lines in both Figures) touch the sidewalls of the PCM elements. The RESET condition is chosen in this way because 880 K is the melting temperature of Ge$_2$Sb$_2$Te$_5$ and we assume the portions of GST enclosed within the 880 K isothermal lines are molten and consequently amorphized after rapid cool-down. Therefore, the shunting paths are eliminated in such a condition, and the PCM devices are converted to the high resistance state. We repeat the same analysis for a mushroom cell geometry, Fig. \ref{fig:Fig_S4} (c) and (d), and observe marginal change in the reset current compared to confined cell geometry.

\newpage
\textbf{Supplementary Note 6 - Transmission Electron Microscopy}
In-situ heating was performed in a FEI Titan TEM at 300 kV equipped with a Gatan OneView camera. A Gatan heating holder (model 652) and Smart Set Hot Stage Controller (model 901) was used to heat the samples incrementally from room temperature by manually setting and ramping the applied current until the desired temperatures was attained. The manually ramping allowed for live continuous acquisition (in-situ movie) while transitioning from 25 to 240 \degree C and 240 to 400 \degree C, in addition to imaging and diffraction at 25, 240, and 400 \degree C where the temperature was held constant within $\pm$5 \degree C. All diffraction patterns were taken using the OneView operating in DP mode to maximize the dynamic range of the camera. Selected-area diffraction patterns were acquired using an aperture collecting from an area 160 nm in diameter. The large collection region of the selected-area aperture relative to the thin-film thickness allowed sampling from both the GST layer and Si substrate providing a self-consistent calibration for GST amorphous ring patterns and crystalline diffraction patterns.

\begin{figure}[H]
\begin{center}
\includegraphics[width=\columnwidth]{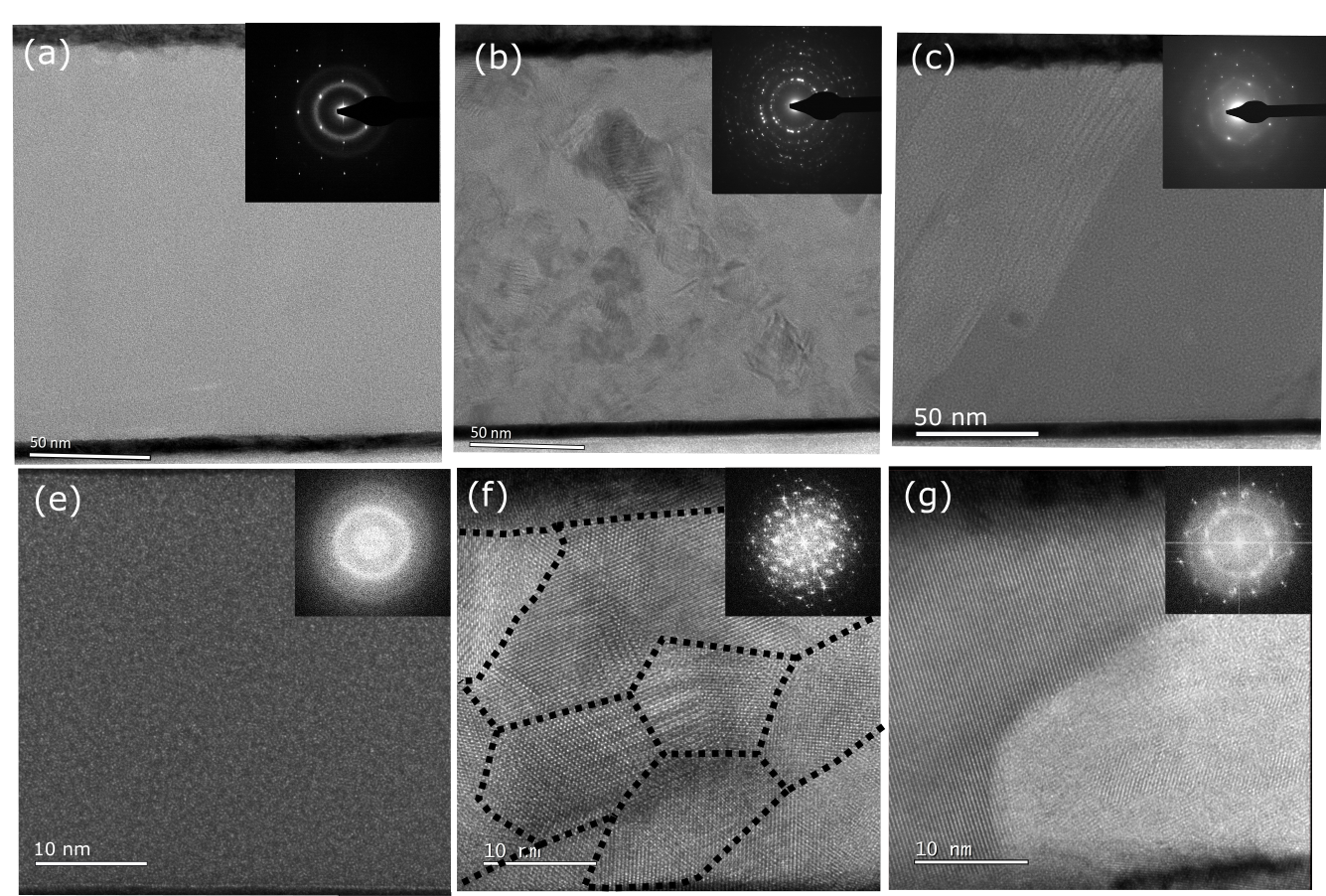}
\caption{TEM images for a-GST, c-GST, and h-GST phases and their corresponding diffraction patterns.
\label{fig:Fig_S10}}
\end{center}
\end{figure}

\begin{figure}[H]
\begin{center}
\includegraphics[width=\columnwidth]{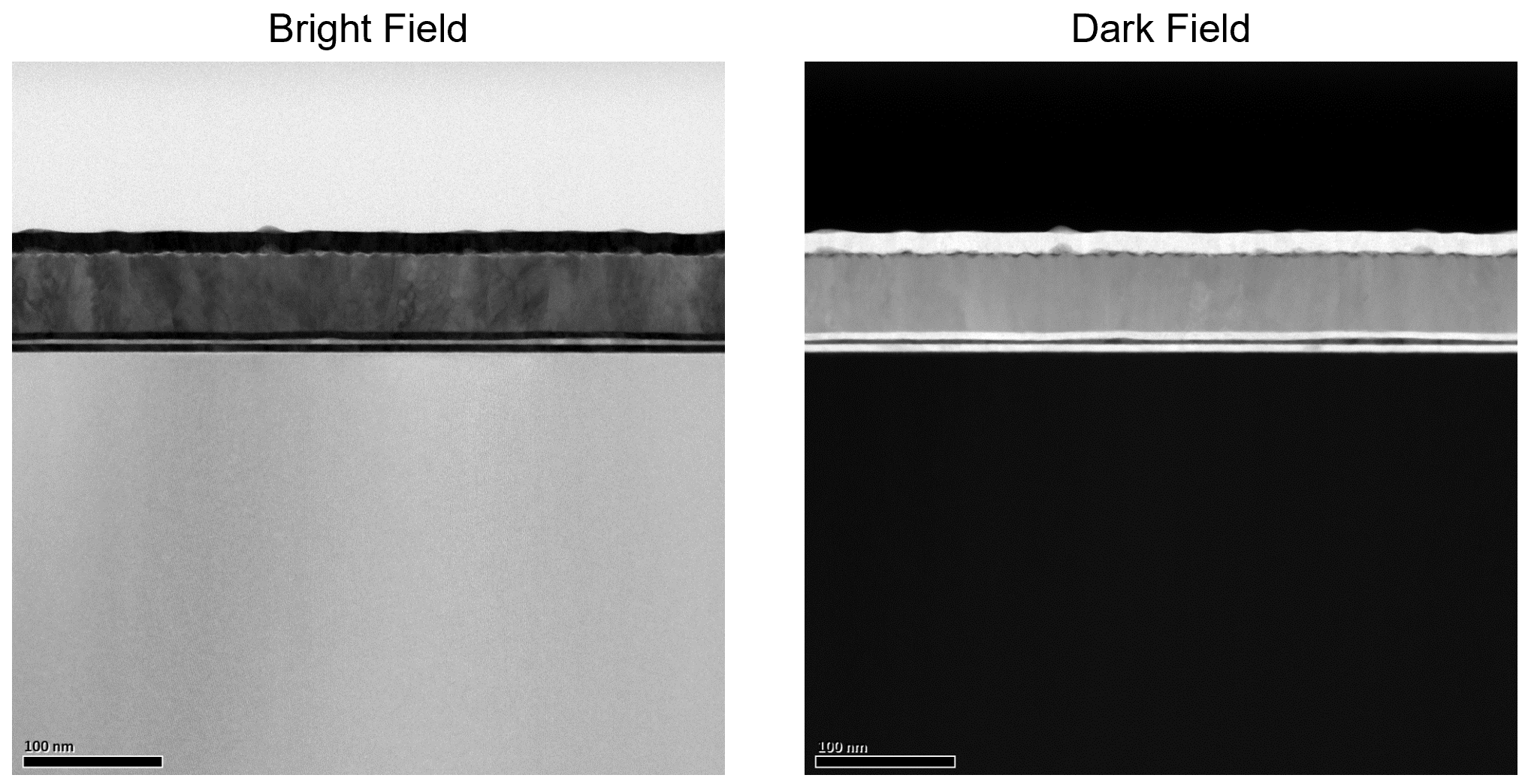}
\caption{STEM images for 5 nm thick a-GST.
\label{fig:Fig_S11}}
\end{center}
\end{figure}

\bibliographystyle{unsrt}
\bibliography{supplement.bbl}